\pdfoutput=1
\documentclass[fleqn,useAMS,usenatbib]{mnras}
\usepackage{graphicx, wasysym, siunitx, subfig}
\usepackage[T1]{fontenc}
\usepackage{aecompl}

\DeclareSIUnit\gauss{G}
\DeclareSIUnit\year{yr}

\title[The Evolving Magnetic Topology of $\tau$ Bo\"otis]{The Evolving Magnetic Topology of $\tau$ Bo\"otis}
\author[M. W. Mengel et al.]{M. W. Mengel,$^{1}$\thanks{E-mail: matthew.mengel@usq.edu.au (MWM)} 
R. Fares,$^{2}$ 
S. C. Marsden,$^{1}$ 
B. D. Carter,$^{1}$ 
S. V. Jeffers,$^{3}$ \and 
P. Petit,$^{4,5}$
J.-F. Donati,$^{4,5}$
C. P. Folsom$^{6,7}$ and 
the BCool Collaboration
\\
$^{1}$Computational Engineering \& Science Research Centre, University of Southern Queensland, Toowoomba, Qld, Australia\\
$^{2}$School of Physics and Astronomy, University of St Andrews, St Andrews KY16 9SS, UK \\
$^{3}$Institut f\"ur Astrophysik, Georg-August-Universit\"at G\"ottingen, Friedrich-Hund-Platz 1, 37077 G\"ottingen, Germany \\
$^{4}$Universit\'e de Toulouse, UPS-OMP, Institut de Recherche en Astrophysique et Plan\'etologie, F-31400 Toulouse, France \\ 
$^{5}$CNRS, Institut de Recherche en Astrophysique et Plan\'etologie, 14 Avenue Edouard Belin, F-31400 Toulouse, France \\
$^{6}$Universit\'e Grenoble Alpes, IPAG, F-38000 Grenoble, France \\
$^{7}$CNRS, IPAG, F-38000 Grenoble, France
}

\begin{document}

\date{Accepted 2016 April 7. Received 2016 April 5; in original form 2015 November 9}

\pagerange{\pageref{firstpage}--\pageref{lastpage}} \pubyear{2015}

\maketitle

\label{firstpage}


\begin{abstract}
We present six epochs of spectropolarimetric observations of the hot-Jupiter-hosting star $\tau$ Bo\"otis that extend the exceptional previous multi-year data set of its large-scale magnetic field.
Our results confirm that the large-scale magnetic field of $\tau$ Bo\"otis varies cyclicly, with the observation of two further magnetic reversals; between December 2013 and May 2014 and between January and March 2015.  We also show that the field evolves in a broadly solar-type manner in contrast to other F-type stars.  We further present new results which indicate that the chromospheric activity cycle and the magnetic activity cycles are related, which would indicate a very rapid magnetic cycle.  As an exemplar of long-term magnetic field evolution, $\tau$ Bo\"otis and this long-term monitoring campaign presents a unique opportunity for studying stellar magnetic cycles.  
\end{abstract}



\section{Introduction}

Magnetic fields of planet-hosting stars are of significant interest given the expected role of the magnetic field in both stellar and planetary system evolution.  $\tau$ Bo\"otis (HR 5185, HD 120136, F7V, age $\sim\SI{1}{\giga\year}$; list of stellar parameters given in Table~\ref{table:stellarParams}) hosts a hot Jupiter with a mass of $\sim 6M_{Jupiter}$ \citep{b28,b27,b26} orbiting at 0.049 AU in approximately 3.31 days \citep{b8,b9,b26} and has been the subject of periodic observation of its magnetic field since 2007.  This unique long-term spectropolarimetric observational series of $\tau$ Bo\"otis has allowed the investigation of the evolution of its magnetic topology over an extended period.

\cite{b2} and \citet{b4,b5} have observed that $\tau$ Bo\"otis exhibits a magnetic cycle including polarity reversals occurring roughly on a yearly timescale.  Studies of younger stars which like $\tau$ Bo\"otis have shallow convective zones \citep{b36,b11,b12}, have not shown magnetic cycles; perhaps because the duration of the cycles are longer than the periods of observation or because their cycles are irregular or chaotic. \cite{b4} speculate that the hot Jupiter with $M_{\mbox{p}} \sim 6M_{Jupiter}$ (compared to the stellar convective envelope with a mass of $\sim 0.5M_{Jupiter}$) may accelerate the stellar magnetic activity cycle by synchronising the outer convective envelope of the star (due to tidal interactions) and enhancing the shear at the tachocline.

$\tau$ Bo\"otis has been a target of interest for those searching for star-planet interaction (SPI). Photometric observations from the \textit{MOST} satellite by \cite{b15} suggested that a persistent active region exists on the star synchronised with the period of the hot Jupiter, but leading the subplanetary longitude by $\sim 68^{\circ}$.  The presence of a persistent active region may suggest star-planet interaction (SPI).  \citet{b29} contends that models rule out that any such ``hot spots'' are due to \textit{magnetic} SPI, and therefore if truly related to the planetary period they must be via some other mechanism.  Given that the orbital period of the planet $\tau$ Bo\"otis~b and the star's rotational period are presumed to be synchronized, observations of the star may not resolve whether or not any rotationally modulated chromospheric features are indicative of SPI or not.

More recently, \citet{b26} analysed spectra of $\tau$ Bo\"otis using the HARPS-N spectrograph. Their study suggests that a high-latitude plage was present near one pole of the star.  While \cite{b26} conclude that ``it is unclear if it is due to SPI or to a corotating active region, or both'', this observation is particularly interesting as it overlaps with one of the epochs presented in this work and is discussed in the conclusion.

We present in this paper a new set of epochs to extend the spectropolarimetric study of $\tau$ Bo\"otis as part of the BCool collaboration\footnotemark on the magnetic fields of cool stars.  Radial magnetic maps from previous epochs of observation have been used as boundary conditions for modelling the wind environment around the star \citep{b44}.  Similarly, work presented here provides these boundary conditions for an ongoing monitoring of the stellar wind of $\tau$ Bo\"otis \citep{b45}.

Investigation of the Ca~\textsc{ii}~H\&K stellar activity proxy of the star is shown in Section~\ref{sec:cahk}.  Modelling of the large-scale magnetic field of the star, including its differential rotation is presented in Section~\ref{sec:mapping}. We draw conclusions in Section~\ref{sec:conclusion}.

\footnotetext{\url{http://bcool.ast.obs-mip.fr}}

\begin{figure}
  \includegraphics[scale=0.40]{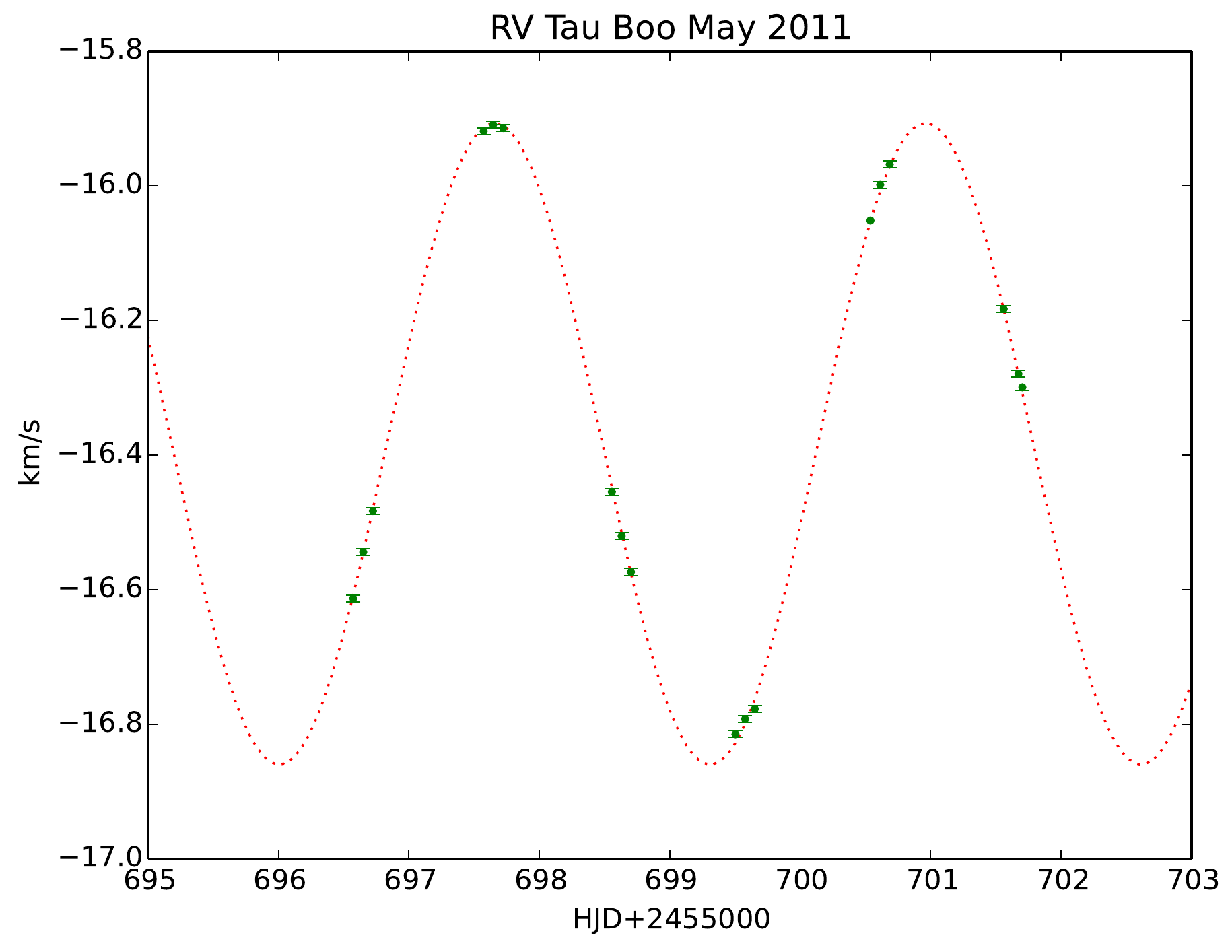}
  \includegraphics[scale=0.40]{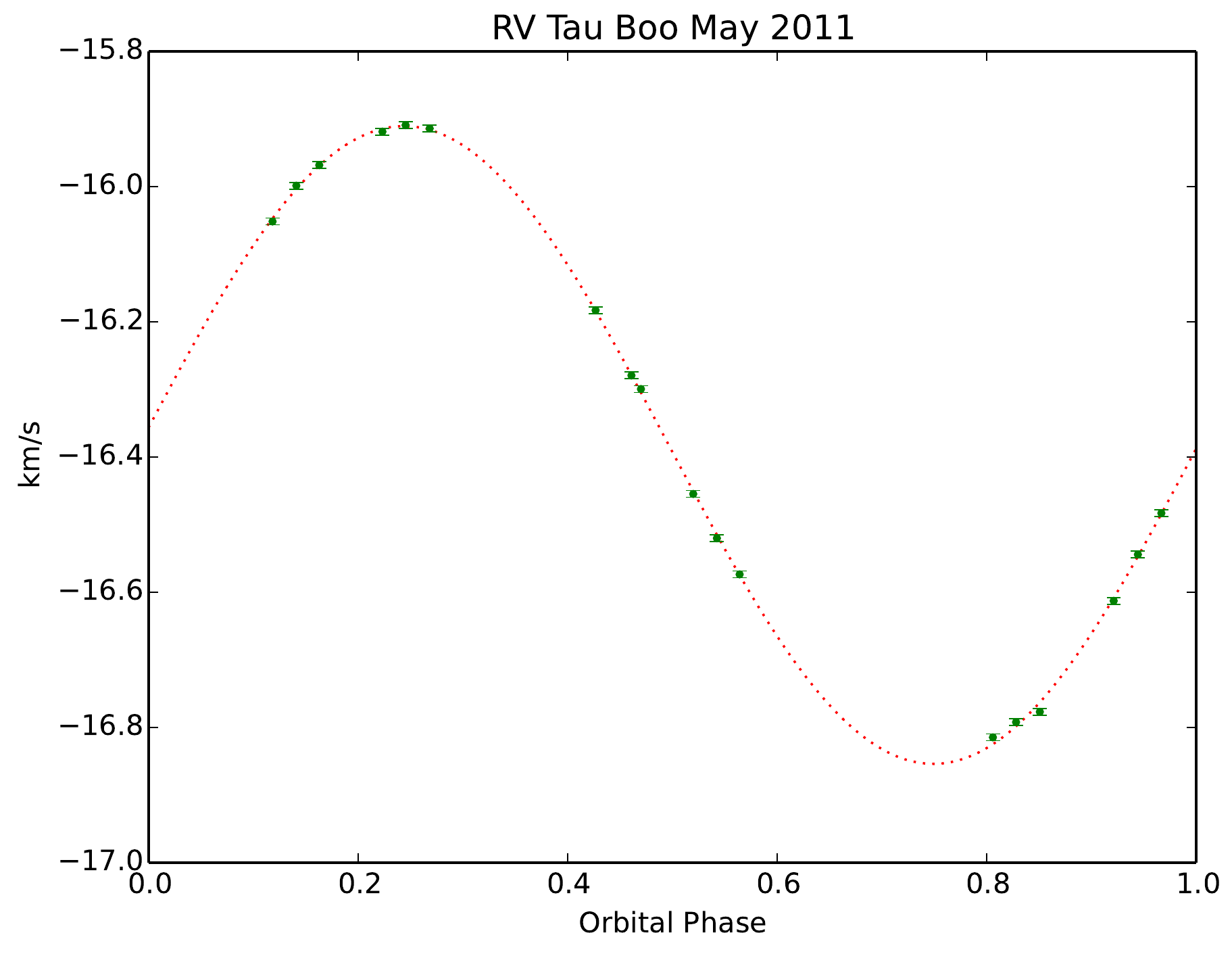}
  \caption{Radial velocity of $\tau$ Bo\"otis, May 2011 derived from HARPSpol data as a function of HJD (upper panel) and as a function of orbital phase (lower panel). The errors in the measurements are  $\sim\SI{5}{\meter\per\second}$, which is the approximate error for HARPSpol in pure spectroscopic mode (\protect\citet{b20} and the accompanying HARPS performance summary\protect\footnotemark).  The fitted semi-amplitude of \SI{476}{\meter\per\second} is close to that in \protect\cite{b8}.}
  \label{fig:RV2011}
\end{figure}

\footnotetext{\url{http://www.eso.org/sci/facilities/lasilla/instruments/harps/inst/performance.html}}

\begin{table}
 \centering
  \caption{Table of stellar parameters for $\tau$ Bo\"otis. (References: 1~=~\protect\citet{b26}; 2~=~\protect\citet{b27}) }
\label{table:stellarParams}
  \begin{tabular}{@{}cccc@{}}
  \hline
   Parameter & & Value & Reference \\
 \hline
T$_{\mbox{eff}}$ & (\SI{}{\kelvin}) & $6399 \pm 45$ & 1 \\
$\log g$ & (\SI{}{\centi\meter\per\second}) & $4.27 \pm 0.06$ & 1  \\
{[}Fe/H{]} & & $0.26 \pm 0.03$  & 1 \\
$v \sin i$ & (\SI{}{\kilo\meter\per\second}) & $14.27 \pm 0.06 $ & 1 \\
Luminosity & ($L_{\odot})$ & $3.06 \pm 0.16$ & 1 \\
Mass & ($M_{\odot})$ & $1.39 \pm 0.25$ & 1 \\
Radius & ($R_{\odot})$ & $1.42 \pm 0.08$ & 1 \\
Age & (\SI{}{\giga\year}) & $0.9 \pm 0.5$ & 1 \\
Inclination & ($^{\circ}$) & $44.5 \pm 1.5$ & 2 \\
\hline
\end{tabular}
\end{table}


\section[]{Observations and Data Processing}
\label{sec:obs}

Stokes V spectropolarimetric data of $\tau$ Bo\"otis were obtained in May 2011 using the HARPS polarimeter (hereafter referred to as HARPSpol) and in 2013, 2014 and 2015 using the NARVAL high-resolution spectropolarimeter.  Each Stokes V spectrum is derived from a sequence of four subexposures taken with the waveplates (HARPSpol)/retarder rhombs (NARVAL) of the polarimeters in different positions \citep{b41,b1}.  The phases of the data are derived using the same orbital ephemeris as that used by \cite{b3}, \cite{b2}, \cite{b4} and \cite{b5}:

\begin{equation}
 T_0 = \mbox{HJD } 2453450.984 + 3.31245E
  \label{eq:ephem}
\end{equation}

with phase 0.0 denoting the first conjunction (i.e. the planet furthest from the observer).

\subsection{Observations with HARPSpol}

HARPSpol \citep{b38} is located at the ESO 3.6-m telescope at La Silla.  HARPSpol has a spectral resolution of around 110000, with spectral coverage from \SIrange{380}{690}{\nano\meter}.  In May 2011, 18 spectra were collected over six nights using HARPSpol, providing good coverage of the complete rotational cycle.  The journal of observations from May 2011 is shown in Table~\ref{table:obsHARPS}.

\subsection{Observations with NARVAL}

NARVAL is attached to the 2-m T\'elescope Bernard Lyot (TBL) at Pic du Midi.  A twin of the CFHT ESPaDOnS instrument \citep{b37}, NARVAL has a spectral coverage of \SIrange{370}{1048}{\nano\meter} with a resolution of approximately 65000.  Information on NARVAL can be found in \citet{b30}. 

From April 23 until May 13 2013 (approximately 21 nights) eight observations were taken using NARVAL.  The coverage of the stellar cycle was incomplete, with two observations close to $\phi_{rot} \sim 0.25$ and the  remaining six giving relatively even coverage of $0.5 < \phi_{rot} < 1$.  Twelve observations were taken using NARVAL from December 4 until December 21 2013 (17 nights).  The coverage of the stellar surface was sparse but relatively complete.  Eleven observations were taken using NARVAL between May 4 and May 18 2014 (14 nights).  In January 2015, seven observations were obtained using NARVAL over 12 nights.  The journal of observations from April 2013 through January 2015 is shown in Table \ref{table:obsNARVAL}.

Beginning in March 2015 a further series of observations of $\tau$ Bo\"otis were obtained. The observations were more sparsely separated and spanned a significant period of time ($\sim\SI{70}{\day}$, $\sim22$ rotations).  Given that this period is considered too long for a single ZDI analysis due to potential for feature evolution, this data set was divided into multiple overlapping epochs and analysed separately.  Thus this observational data is presented in its own journal of observations (Table~\ref{table:obsNARVAL2015}).

\subsection{Data Reduction}

Data from NARVAL was automatically reduced using the \textsc{libre-esprit} software package.  A specifically modified version of \textsc{libre-esprit} was used by JFD to reduce the data from HARPSpol.  \textsc{libre-esprit} produces Stokes I (unpolarized) and Stokes V (circularly polarized) spectra, in addition to a null (N) spectrum which is used to determine the authenticity of a detected polarization signal \citep{b1}.

\subsubsection{Least Squares Deconvolution (LSD)}

As Zeeman signatures are typically smaller than the noise level within a single spectral line, they are difficult to detect, especially for solar-type stars such as $\tau$ Bo\"otis. The Least Squares Deconvolution (LSD) technique is applied to improve the S/N of the data by combining the information provided by many spectral lines  \citep{b1}.  The line mask used to perform the deconvolution is the same as described in \cite{b2} and \cite{b4}, using a Kurucz model atmosphere with solar abundances, an effective temperature of \SI{6250}{\kelvin} and $\log g$ of \SI{4.0}{\centi\meter\per\second\squared}, including most strong lines in the optical domain (central depths $>$$\sim$ 40 percent of the local continuum before macroturbulent or rotational broadening) but excluding the strongest, broadest features such as Balmer lines.  This results in each decovolution utilizing 3000-4000 lines, depending on the particular spectral coverage of the instrument.

In the journals of observations (Tables~\ref{table:obsHARPS}, \ref{table:obsNARVAL}, \ref{table:obsNARVAL2015}), a definite detection (D) in the LSD Stokes V profile is defined as a false alarm probability ($fap$) of less than $10^{-5}$.  A marginal detection (M) has a false alarm probability greater than $10^{-5}$ but less than $10^{-3}$ \citep{b1}.

\subsubsection{Radial Velocity}

Due to the presence of the hot Jupiter in orbit around $\tau$ Bo\"otis, the radial velocity (RV) of the star varies from observation to observation.  The best-fit semi-amplitude of \SI{476}{\meter\per\second} we derive is close to that found by \protect\cite{b8} and in good agreement with the expected phasing using the orbital ephemeris of \cite{b3}  (See Fig.~\ref{fig:RV2011} for May 2011 observations).  The spectra are automatically corrected by the \textsc{libre-esprit} software for RV variations due to the motion of the Earth. The RV of the star due to the orbital motion of the system is derived by fitting the Stokes I LSD profile of each observation with a Gaussian and determining the centre of the profile.  The spectra are then corrected for the RV due to the system's orbital motion.




\section[]{Ca \textsc{ii} H \& K Activity Proxy}
\label{sec:cahk}

\begin{figure*}
  \includegraphics[scale=0.4]{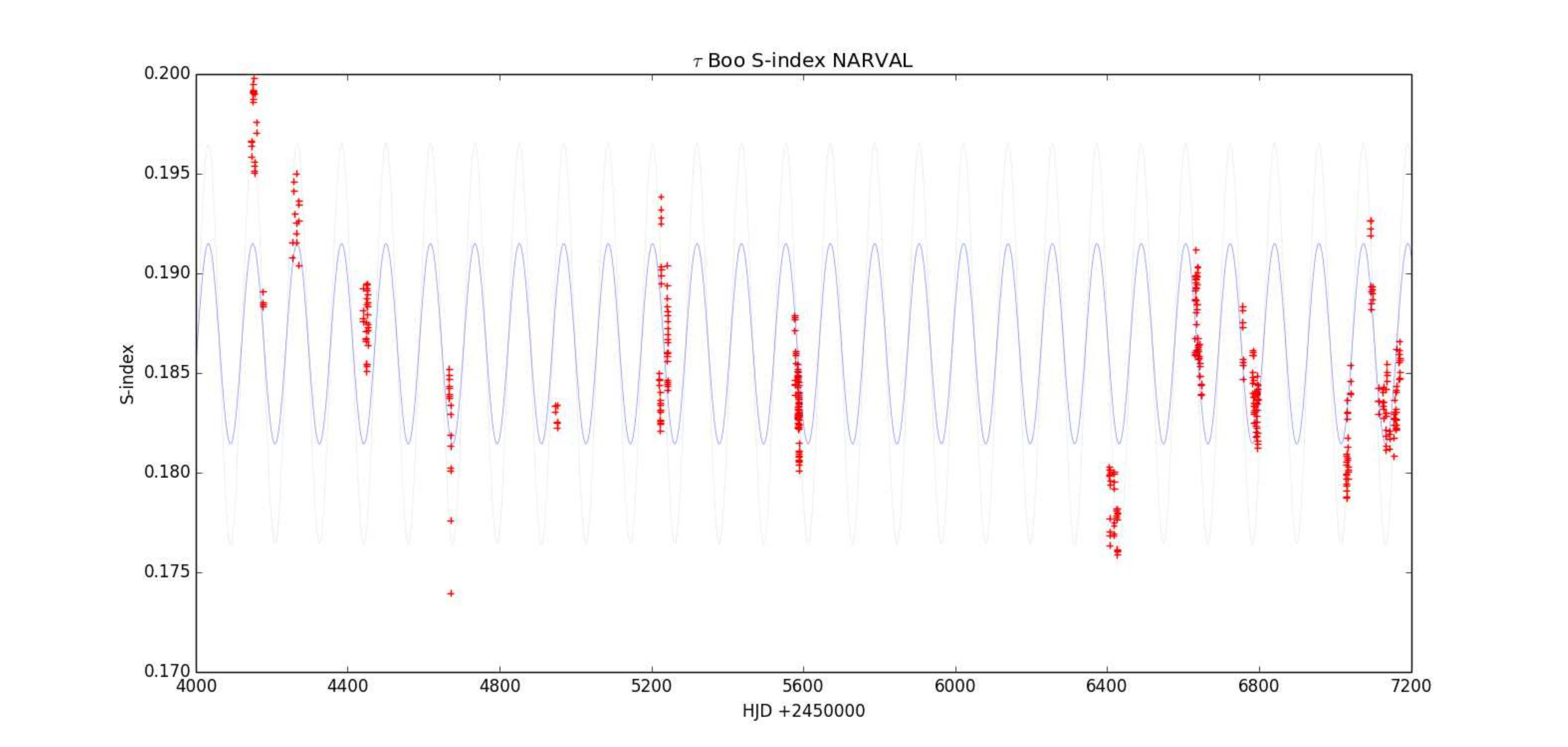}
    \caption{Ca II HK S-indices for $\tau$ Bo\"otis.  NARVAL observations are shown in red.  A least-squares fit of a sinusoid to the unweighted data (blue line) yields a period of $\sim\SI{117}{\day}$, which corresponds closely to the $\sim\SI{116}{\day}$ period reported by \protect\cite{b21,b25}.}
  \label{fig:cahk}
\end{figure*}

The emission in the cores of the Ca \textsc{ii} H \& K lines is one of the most widely used proxies for stellar chromospheric activity.  The S-index for each observation of $\tau$ Bo\"otis was calculated using the method of  \cite{b19} and utilizing the coefficients for NARVAL derived by \cite{b18} for the equation:

\begin{equation}
\mbox{S-index} = \mbox{Ca}_{\mbox{\textsc{hk}}}\mbox{-index} = {{aF_H + bF_K} \over { cF_{R_{HK}} + dF_{V_{HK}}}} + e
  \label{eq:sindex}
\end{equation}

where $F_H$ and $F_K$ are the fluxes in a 2.18\r{A} triangular bandpasses centred on the cores of the Ca \textsc{ii} H \& K lines, $F_{R_{HK}}$ and $F_{V_{HK}}$ are rectangular 20\r{A} bandpasses centred on the continuum at 3901.07\r{A} and 4001.07\r{A} \citep[Fig.~1]{b19}.  These coefficients for the NARVAL instrument are shown in Table~\ref{table:scoeffs}.

\begin{table}
 \centering
  \caption{Table of coefficients for Equation~\protect\ref{eq:sindex} as calculated by \protect\cite{b18} for the NARVAL instrument.}
\label{table:scoeffs}
  \begin{tabular}{@{}ccc@{}}
  \hline
   Coefficient & NARVAL \\
 \hline
$a$ & 12.873 \\
$b$ & 2.502 \\
$c$ & 8.877 \\
$d$ & 4.271 \\
$e$ & $1.183 \times 10^{-3}$ \\
\hline
\end{tabular}
\end{table}

Following the methodology of \citet{b18}, overlapping orders were removed from the reduced spectra of $\tau$ Bo\"otis and adjusted for the radial velocity for the observation.  In addition to the newly observed epochs, all prior observations of $\tau$ Bo\"otis with the NARVAL instrument were retrieved from the Polarbase\footnote{\url{http://polarbase.irap.omp.eu/}} \citep{b17} database, and we calculated values of the S-index for each normalised individual exposure of $\tau$ Bo\"otis, making four data points for each spectropolarimetric sequence.  These are shown in Figure~\ref{fig:cahk}.  As noted in \citet{b4}, $\tau$ Bo\"otis exhibits intrinsic variability through each night and night-to-night, and this can be observed in the data.  Using a least-squares fit of a sinusoid to the unweighted S-index data we find a longer-term variability in chromospheric activity of $\sim\SI{117}{\day}$.  This result corresponds well to the $\sim\SI{116}{\day}$ period reported by \protect\citet{b21} and \protect\citet{b25} from the Mount Wilson HK project \citep{b42}. 

There also appears to be a longer-term trend apparent in the data.  \citet{b21} reports a low-amplitude \SI{11.6}{\year} activity cycle, however our data set is too short to reliably perform a fit of that duration.  In attempting to fit multiple periods, various results converged depending upon initial conditions.  There are several fits of equally good quality with different periodicities from approximately $\sim\SI{300}{\day}$ to several thousand days and given the large gaps in the observational records, these periods are uncertain.  Further observational epochs taken at shorter intervals are be required to accurately characterise the cyclic behaviour of the Ca \textsc{ii} H \& K activity proxy of $\tau$ Bo\"otis and if any relationship to other observed cycles exists.

\section[]{Magnetic Mapping}
\label{sec:mapping}

\subsection{Model Description}
\label{sec:modeldesc}

Zeeman Doppler Imaging (ZDI) is used to reconstruct maps of the magnetic topology of $\tau$ Bo\"otis from the observed Stokes V signatures (Fig.~\ref{fig:profiles}). The process uses the principles of maximum entropy image reconstruction to produce the configuration of the large-scale magnetic field containing the minimum information required to produce the observed magnetic signatures.  The code used is that described in \citet{b10} wherein spherical-harmonic expansions are used to describe the field configuration with respect to its poloidal and toroidal components.  An advantage of this method is that the coefficients of the spherical harmonics can be used to calculate the energy contained in, for example, axissymetric and non-axissymetric modes, and to determine relative contributions of dipolar, quadrupolar and higher-order components.

The stellar surface is divided into units of similar projected area and the contribution of each unit area to the Stokes V profile (based on field strength, orientation, surface location and motion) is calculated.  The process continues iteratively wherein profiles are reconstructed and compared to the observed profiles until a match within the desired error is reached (typically a unit reduced $\chi^{2}$; i.e. $\chi_{r}^{2} \sim 1$). 

The local Stokes I profile is modelled by a Gaussian (FWHM of \SI{11}{\kilo\meter\per\second}), while the local Stokes V is calculated assuming the weak field approximation \citep{b1}:

\begin{equation}
 V \propto g B_{los} {dI \over dv}
  \label{eq:wkfld}
\end{equation}

where $B_{los}$ is the local line-of-sight component of the magnetic field and $g$ is the mean land\'e factor.

\subsection{Differential Rotation}
\label{sec:dr}

\subsubsection{Method}
\label{sec:dr:method}

When a star is differentially rotating, the signatures produced by magnetic regions will repeat from rotational cycle to rotational cycle but with differences resulting from shifts in the relative location of the regions due to the differential rotation.  We consider that the rotation will follow a simplified solar-type law:

\begin{equation}
 \Omega(\theta) = \Omega_{eq} - d\Omega \sin^{2} \theta
  \label{eq:dr}
\end{equation}

where $\Omega(\theta)$ is the rotation rate of the star at latitude $\theta$ in \SI{}{\radian\per\day}, $\Omega_{eq}$ is the rotation rate at the equator and $d\Omega$ is the rotational shear between the equator and the poles. 

Applying the method described by \cite{b39}, \citet{b22}, \citet{b6} and \citet{b7} we construct a magnetic image containing a given information content for each pair of ($\Omega_{eq}$, $d\Omega$) and choose the pair of parameters which produces the best fit to the data (i.e. the smallest $\chi_{r}^{2}$).  An example for the December 2013 data set is shown in Figure~\ref{fig:drms}(a).  Figure~\ref{fig:drms}(b) shows how we derive $\sim$1-$\sigma$ variation bars as a measure of uncertainty by varying various stellar parameters ($v \sin i \pm \SI{1}{\kilo\meter\per\second}$; inclination $\pm \ang{10}$; target reconstructed average magnetic field $B_{mod} \pm \sim 10\%$) and calculating the extreme variations.

\citet{b2} measured the differential rotation of $\tau$ Bo\"otis utilizing the Stokes V data, equal to $\Omega_{eq} =  \SI[separate-uncertainty = true]{2.10(4)}{\radian\per\day}$ and $d\Omega = \SI[separate-uncertainty = true]{0.50(12)}{\radian\per\day}$.  However, subsequent observations have yielded either different values of $\Omega_{eq}$ and $d\Omega$ (e.g. January 2008, $d\Omega = 0.28 \pm 0.10$; \citet{b4}) or no measurement has been possible \citep{b5}.

\subsubsection{Results}
\label{sec:dr:results}

The May 2011 HARPSpol data produced a well-defined parabaloid, albeit with a larger error compared to other runs.  The HARPSpol data is of a lower S/N than that from NARVAL, however the May 2011 phase coverage is superior to the other runs.  The derived parameters of $\Omega_{eq} = 2.03^{+0.05}_{-0.05}$ and $d\Omega = 0.42^{+0.11}_{-0.11}$~\SI{}{\radian\per\day} is consistent with those of \citet{b2}.

\begin{figure*}
\begin{tabular}{cc}
 \includegraphics[width=75mm]{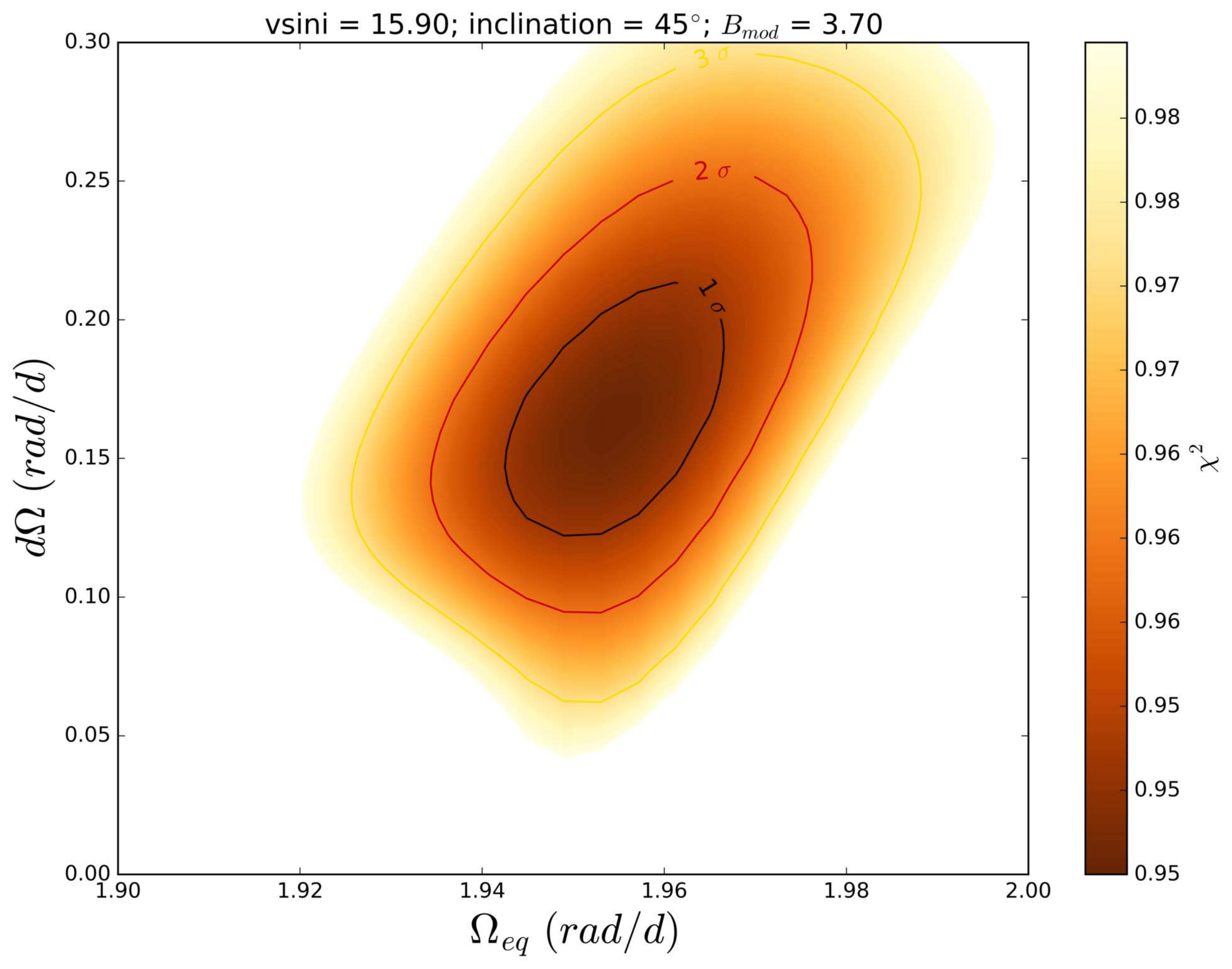}  &   \includegraphics[width=80mm]{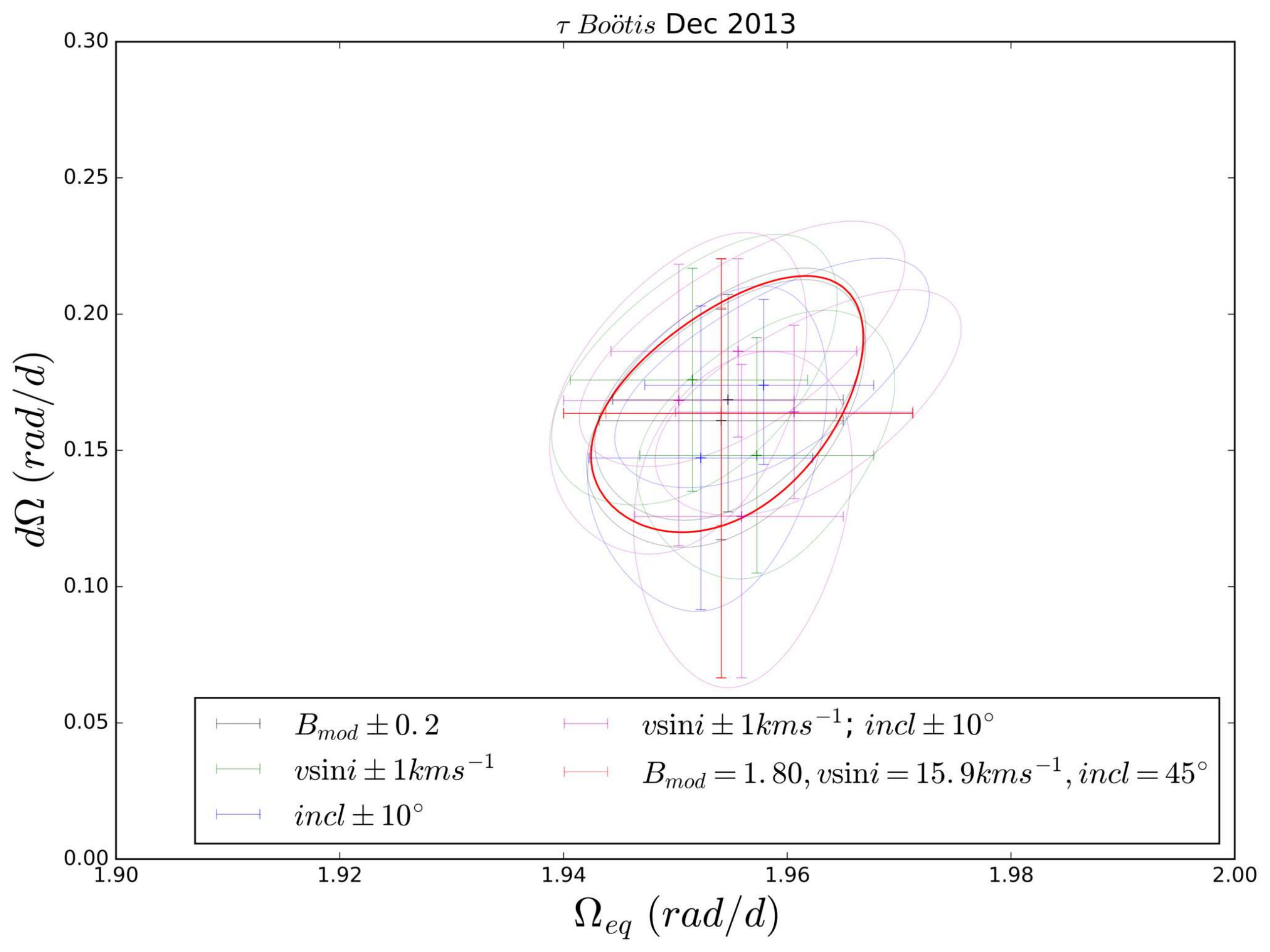} \\
(a) $\chi_{r}^{2}$ as a function of $\Omega_{eq}$ and $d\Omega$ & (b) Second-order polynomial fit to 1-$\sigma$ contour in (a) in red \\
 & and fits to variations of stellar parameters \\[6pt]
\end{tabular}
\caption{Differential rotation of $\tau$ Bo\"otis for December 2013. Panel (a) shows the variation of $\chi_{r}^{2}$ as a function of $\Omega_{eq}$ and $d\Omega$ for the selected stellar parameters ($v \sin i = \SI{15.9}{\kilo\meter\per\second}$, inclination \ang{45}) and a target $B_{mod}$ appropriate for the desired $\chi_{r}^{2}$ used to reconstruct the magnetic maps of this epoch.  Panel (b) shows second-order polynomial fits for the 1-$\sigma$ threshold parabaloids surrounding the minimum $\chi_{r}^{2}$ where the red paraboloid corresponds to the plot in panel (a).  Other paraboloids represent fits to the corresponding DR calculations varying the target $B_{mod} (\pm \sim 10\%, \mbox{black})$, $v \sin i (\pm \SI{1}{\kilo\meter\per\second}$) and stellar inclination ($\pm \ang{10}$) independently (green and blue) and $v \sin i$ and inclination angle together (magenta).  The overall variation bars are used to derive the uncertainty in the measured DR parameters.}
  \label{fig:drms}
\end{figure*}

For the NARVAL data from April/May 2013, we measure a significant error bar.  The maps for April-May 2013 show that there are only two observations of the star consistent with phase $\sim$0.25 ($\phi_{rot}$ of -2.749 and +3.262), and these are separated by $\sim$6 full rotations ($\sim\SI{20}{\day}$).  Removing these observations results in a smaller error and a lower value of $d\Omega$. However, the DR measurement used in our mapping and reported in Table~\ref{table:dr} is made with all observations present, hence a much larger error is calculated.  The derived parameters for April/May 2013 were $\Omega_{eq} = 2.05^{+0.04}_{-0.04}$ and $d\Omega = 0.38^{+0.18}_{-0.19}~\SI{}{\radian\per\day}$.

With a better coverage of the stellar rotational cycle in December 2013, a clear parabloid was again generated (Fig.~\ref{fig:drms}(a)).  The measured differential rotation for December 2013 corresponded to parameters of $\Omega_{eq} = 1.95^{+0.01}_{-0.01}~\SI{}{\radian\per\day}$ and $d\Omega = 0.16^{+0.04}_{-0.04}~\SI{}{\radian\per\day}$.

The data for May 2014 once more produced a clear parabaloid with the parameters $\Omega_{eq} = 1.99^{+0.01}_{-0.01}~\SI{}{\radian\per\day}$ and $d\Omega = 0.10^{+0.04}_{-0.04}~\SI{}{\radian\per\day}$.  Finally, the derived parameters for January 2015 were $\Omega_{eq} = 1.98^{+0.03}_{-0.03}~\SI{}{\radian\per\day}$ and $d\Omega = 0.15^{+0.15}_{-0.16}~\SI{}{\radian\per\day}$.

These values are summarised in Table~\ref{table:dr}.

\subsubsection{A discussion of variable Differential Rotation measurement}
\label{sec:dr:interpret}

Table~\ref{table:dr} makes clear that the technique used to determine differential rotation yields a wide variation in $d\Omega$, but is consistent with solar-like differential rotation with the equator rotating faster than the poles.  $\Omega_{eq}$ is much more constrained within a range from $\sim$\SIrange{1.95}{2.05}{\radian\per\day}, ignoring the error.  As simulations performed by \citet{b22} show, factors including the phase coverage and observational cadence can have significant effects on the measured differential rotation parameters.

Notably, apart from the May 2011 HARPSpol observation, at best the observations of $\tau$ Bo\"otis have had a single observation per night (approximately one $4\times\SI{600}{\second}$ observation every one-third of a rotation) over several full stellar rotations.  As the technique used to determine differential rotation is a $\chi^{2}$-landscape technique and dependent upon the regular re-observation of features, it is possible that the significant variation in the measured parameters are due to observational biases.  The weakness of the field of $\tau$ Bo\"otis and the difficulty in observing Zeeman signatures on the star potentially exacerbates this effect.  Attempting the $\chi^{2}$ minimisation method with lower values of target $B_{mod}$ with various $\tau$ Bo\"otis data sets, (corresponding to higher $\chi^{2}_{r}$ fits) thus providing less information for the technique produces a systemic decrease in the calculated values of $\Omega_{eq}$ and $d\Omega$. 

This supposition would need to be confirmed by simulations and it should be noted that this does not rule out the existence of small or significant actual variation in differential rotation of magnetic features on $\tau$ Bo\"otis; simply that we cannot draw any such conclusion from these measurements.  A paper is in preparation (Mengel et. al.) examining the effect of phase coverage on measuring DR parameters on slow rotating weak-field stars and how we may determine optimal observational cadences and periods for this type of target.

Finally, it is to be noted that varying $d\Omega$ will slightly distort the shape of the features on the maps but does not appear to have a significant effect in the determination of the general magnetic field properties (Table~\ref{table:results}).  Consequently to be internally consistent in applying the maximum-entropy, minimum information mapping technique, these measured values for $d\Omega$ and $\Omega_{eq}$ which give the minimum information solution have been utilised in the magnetic mapping in this work.

\subsubsection{Stokes I Differential Rotation - Fourier Transform Method}
\label{sec:dr:reiner}

An alternative measurement of the differential rotation can be made using the Fourier Transform (FT) method described by \citet{b24}.  

\begin{figure}
 \includegraphics[width=\columnwidth]{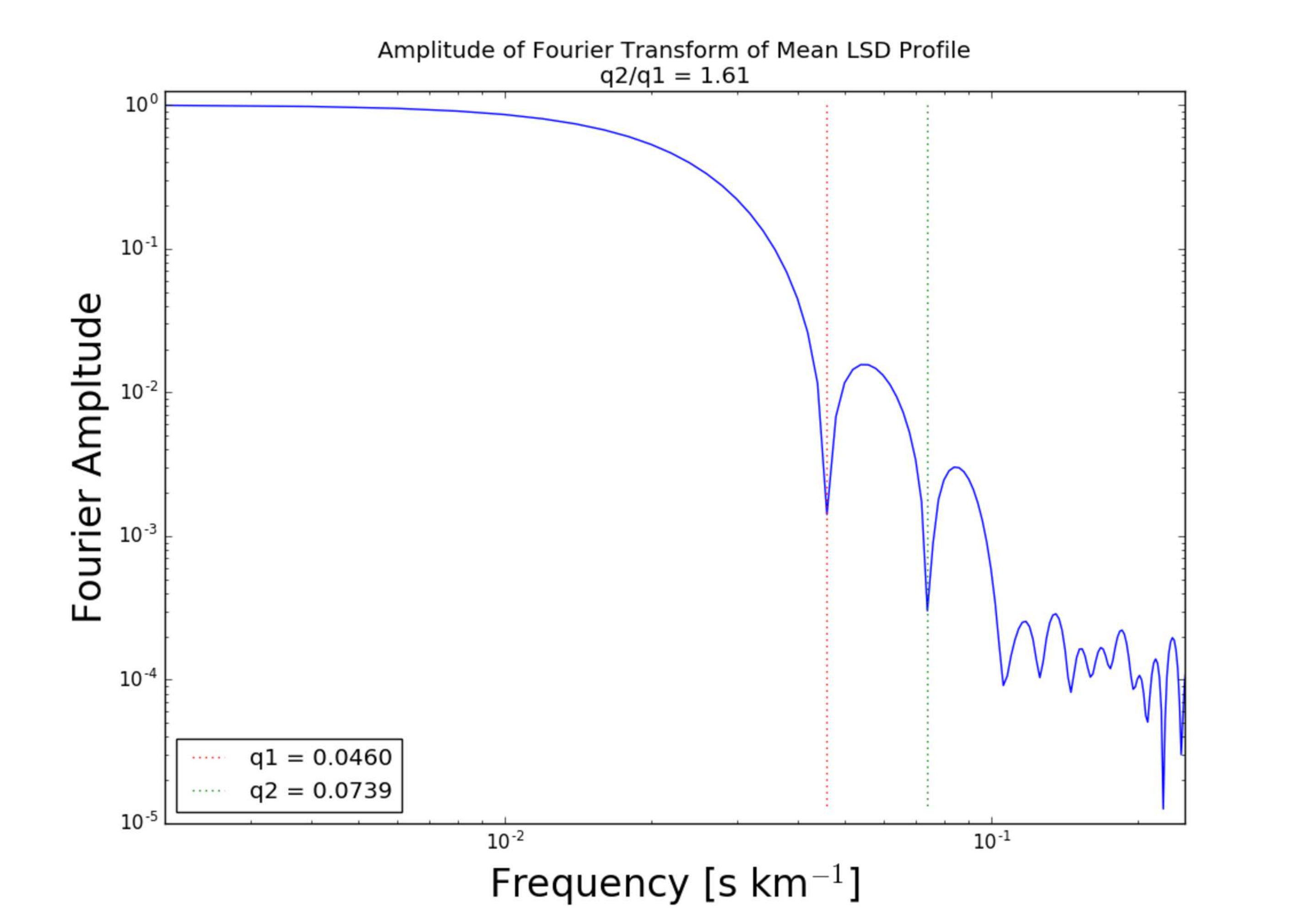}\\
\caption{Amplitude of the Fourier transform of the averaged LSD Stokes I profile for May 2014.  The first two zeros ($q_1, q_2$) are shown.  The ratio $q_2 / q_1$ can be used to derive $\alpha$ ($d\Omega/\Omega_{eq}$) as described in the text (Section~\protect\ref{sec:dr:reiner}).}
  \label{fig:drfft}
\end{figure}

An averaged Stokes I profile was created from the NARVAL and HARPSpol data for each observational epoch.  Using the method of \citet{b24}, we obtained the ratio of the first two zeros of the Fourier transformed average Stokes I line profile ($q_2 / q_1$).  An example of this is shown for the May 2014 epoch in Fig.~\ref{fig:drfft}.  For each epoch, we determined that $q_2 / q_1 = 1.61 \pm 0.07$.  From Equation 5 in \citet{b24}, using an inclination angle of \ang{45} we thus calculate $\alpha$ ($d\Omega/\Omega_{eq}$) of $\approx 0.17$.

Using the various measured values of $\Omega_{eq}$ from Table~\ref{table:dr}, this produces a value for $d\Omega$ of between \SI{0.34}{\radian\per\day} and \SI{0.36}{\radian\per\day}, which is in good agreement with the value found by \citet{b23}, \citet{b3}, and \citet{b26} for $\tau$ Bo\"otis.

It is noted that this technique is quite sensitive to the derived value of $q_2 / q_1$.  Thus using $q_2 / q_1 = 1.61 \pm 0.07$ would yield an uncertainty in $d\Omega$ of \SI{0.14}{\radian\per\day}.  Again, this uncertainty is in agreement with the measurements of the differential rotation of $\tau$ Bo\"otis by \citet{b23}.

Yielding a value of $d\Omega = \SI[separate-uncertainty = true, multi-part-units=single]{0.35 \pm 0.14}{\radian\per\day}$, the Fourier transform method is broadly consistent with the values derived by the $\chi^{2}$-landscape method taking into account the errors calculated for both methods.  Despite the consistency of the two methods described here, the differential rotation measurement must be viewed with caution due to the uncertainties and assumptions inherent in each method.

\begin{table}
 \centering
  \caption{Summary of differential rotation parameters for $\tau$ Bo\"otis as measured by the $\chi^{2}$ minimisation method.  For comparison, the FT method (Sec.~\protect\ref{sec:dr:reiner}, assuming $\Omega_{eq} = \SI{2.00}{\radian\per\day}$) yields $d\Omega = \SI[separate-uncertainty = true, multi-part-units=single]{0.35 \pm 0.14}{\radian\per\day}$.  The values presented below are used in the magnetic mapping.}
\label{table:dr}
  \begin{tabular}{@{}lcc@{}}
  \hline
   Epoch & $\Omega_{eq}$ & $d\Omega$  \\
        & \SI{}{\radian\per\day}  & \SI{}{\radian\per\day} \\
 \hline \\[-1.5ex]
2011 May & $2.03^{+0.05}_{-0.05}$ & $0.42^{+0.11}_{-0.11}$ \\[3pt]
2013 April/May & $2.05^{+0.04}_{-0.04}$ & $0.38^{+0.18}_{-0.19}$ \\[3pt]
2013 December & $1.95^{+0.01}_{-0.01}$ & $0.16^{+0.04}_{-0.04}$ \\[3pt]
2014 May & $1.99^{+0.01}_{-0.01}$ & $0.10^{+0.04}_{-0.04}$ \\[3pt]
2015 January & $1.98^{+0.03}_{-0.03}$ & $0.15^{+0.15}_{-0.16}$ \\[3pt]
\hline
\end{tabular}
\end{table}

\subsection{Magnetic Mapping}
\label{sec:magmap}

\subsubsection{Stellar and Model Parameters}
\label{sec:magmap:param}

Utilising the measured differential rotation parameters, a $\chi^{2}$ minimization process was used to determine the optimum angle of inclination and $v\sin i$.  The values derived were close to those used in previously published works \citep{b4,b5} so an inclination of \ang{45}  and $v \sin i$  of \SI{15.9}{\kilo\meter\per\second} as used in \citet{b5} were chosen and applied to all data sets. (It is noted that this value of $v \sin i$ varies by $\sim10\%$ from the latest published value shown in Table~\ref{table:stellarParams}, however the difference in the models was insignificant using either value, thus we use the derived value for consistency with previous works).

Once the stellar parameters and differential rotation values were chosen, maps were generated for different values of $\chi_{r}^{2}$.

The reconstructed average magnetic field (B$_{mod}$) begins to rise significantly with a lower target $\chi_{r}^{2}$ as the process begins fitting inappropriately to the noise inherent in the signatures, this requires the target $\chi_{r}^{2}$ to be chosen with care.  This effect is much more pronounced in the HARPSpol data than for the NARVAL data and it would thus be inappropriate to utilise the same $\chi_{r}^{2}$ for the May 2011 data.  Thus a $\chi_{r}^{2}$ was chosen to be 0.95 for the NARVAL data, and 1.10 for the HARPSpol data.  A potential drawback of this is that the comparison of the absolute magnitude of the field strength may no longer be appropriate for the data sets derived from the different instruments or between data sets using the same instrument with significantly different S/N levels and thus target $\chi^{2}_{r}$.

The spherical harmonic expansions for the magnetic field were calculated using $\ell = 8$ as little improvement was obtained using $\ell > 8$, there being near to zero energy in higher order harmonics.  This is consistent with previous papers  dealing with $\tau$ Bo\"otis and adequate given there are $\sim 9$ spatial resolution elements around the equator as per equation \ref{eq:harmonics} \citep[Eq. 3]{b40}.

\begin{equation}
 { { 2 \pi v_e \sin i } \over {\mbox{FWHM}}}= { { 2 \pi (\SI{15.9}{\kilo\meter\per\second}) } \over { \SI{11}{\kilo\meter\per\second} } } \sim 9
  \label{eq:harmonics}
\end{equation}

The amount of energy in the various harmonics can be used to describe the configuration of the magnetic field \citep{b10}.  These results, shown in Table~\ref{table:results} consist of three calculations.  First, the amount of the magnetic energy stored in the toroidal component of the field is found from the $\gamma_{\ell,m}$ term of the complex coefficients of the spherical harmonics. Second, the amount of magnetic energy in the poloidal component ($\alpha_{\ell,m}$ and $\beta_{\ell,m}$) which is axisymmetric (i.e. symmetric about the axis of rotation) is found from the from the $\alpha_{\ell,m}$ and $\beta_{\ell,m}$ coefficients of the spherical harmonics where $m=0$.  (Other papers, such as \citet{b4,b5} use $m < \ell/2$ rather than $m=0$. We consider $m=0$ to be more mathematically correct, however the difference in practice is quite small so direct comparisons between epochs are possible.  Where appropriate this is noted in the text).  Finally, the amount of magnetic energy in the poloidal component of the magnetic field consisting of dipolar plus quadrupolar components is calculated from  $\alpha_{\ell,m}$ and $\beta_{\ell,m}$ coefficients where $\ell \leq 2$.

As in the differential rotation measurements, an indication of the variation in these values is found by varying the differential rotation measurement by the error found for each epoch (holding inclination and $v \sin i$ constant), then varying the stellar parameters (inclination $\pm \ang{10}$; $v \sin i \pm \SI{1.0}{\kilo\meter\per\second}$) while holding $d\Omega$ and $\Omega_{eq}$ constant.  It is noted that the variation of stellar parameters produces variation in field components in a consistent way across the epochs, and the variations, although sometimes large relative to the values are of the same order of magnitude across the epochs.  The amount of toroidal field has the smallest variation due to parameter variation.  The amount of energy in axissymmetric and $\ell \leq 2$ modes in the poloidal field exhibit large variability with chosen stellar parameters, albeit with the dominant parameter affecting this variation being the chosen angle of inclination.  Where there is a large uncertainty in the measured ($d\Omega$, $\Omega_{eq}$), such as in the May 2011 and January 2015 epochs, the combination of parameters can create significant variation.

As a derived inclination angle for $\tau$ Bo\"otis which is very close to our chosen value with a low level of uncertainty is known (cf. $\ang{44.5}\pm\ang{1.5}$ \citep{b27} vs \ang{45}), the variation of inclination in our process by $\pm \ang{10}$ is somewhat conservative and thus our variation values may be overestimated.  Consequently, we have a reasonable confidence that epoch-to-epoch changes in the field topology are real irrespective of the sometimes large and conservative variation measurements.

As with determining differential rotation, phase coverage must be adequate to successfully reconstruct the features on the stellar surface.  Features, and thus energy in the harmonics may be missed if parts of the stellar surface are not observed.  It is noted that the cadence observations of $\tau$ Bo\"otis in observations after 2011 is lower than previous epochs.  Consequently fewer phases are used in most of the epochs in this work compared to those of \citet{b2} and \citet{b4,b5}.  Reconstruction of maps in this work with partial, sampled data sets shows that the latitude information of features and the overall field configurations seem to be conserved, albeit with a reduction in the overall energy seen in the harmonics.  These effects for a star with a weak magnetic field was discussed briefly by \citet{b43} and is being further investigated in a forthcoming paper from Mengel et. al.

\begin{figure}
\begin{center}
\includegraphics[scale=0.35]{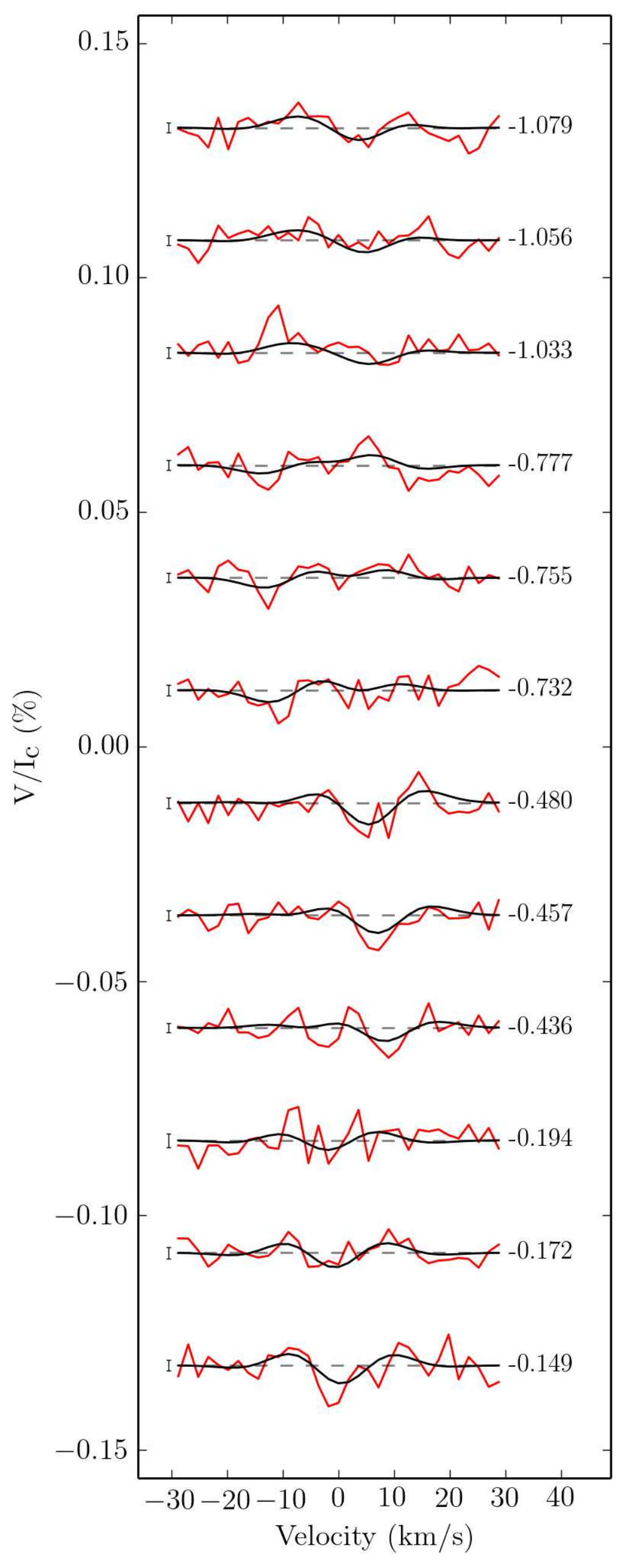}
\includegraphics[scale=0.35]{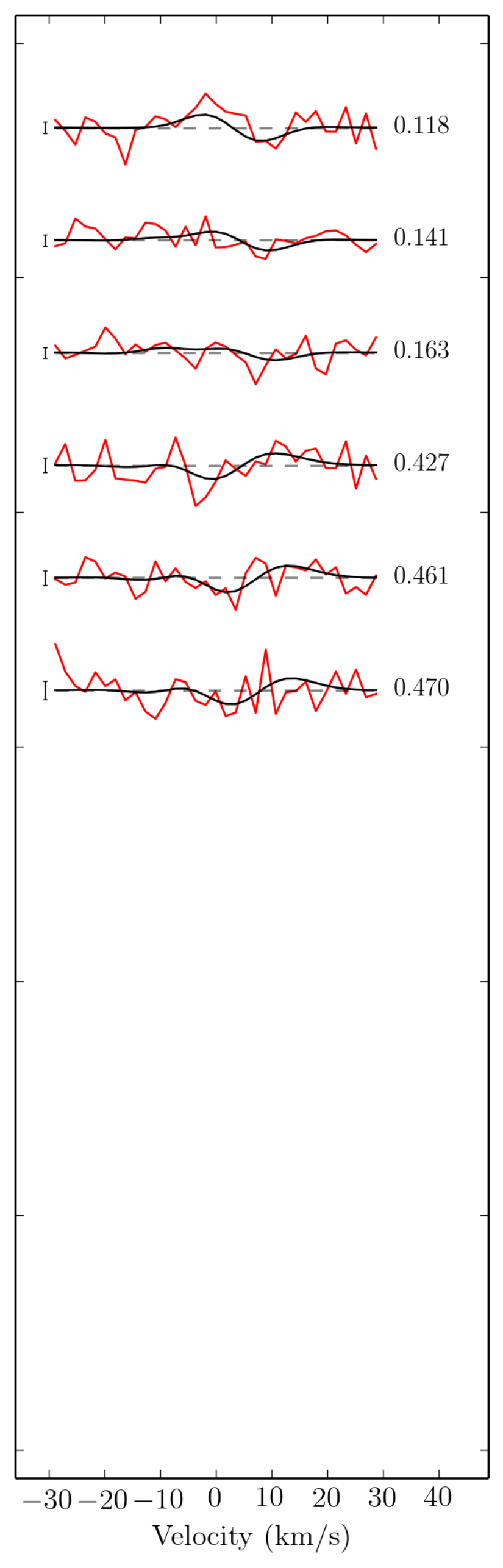}
\end{center}
  \caption{Circular polarization profiles of $\tau$ Boo for HARPSpol observations,  May 2011.  The observed profiles are shown in red, while synthetic profiles are shown in black. On the left of each profile we show a $\pm1\sigma$ error bar. The rotational cycle of each observation is indicated on the right of each profile.  All images of profile fits such as this are on the same scale across this work.}
  \label{fig:profilesHARPS}
\end{figure}

\begin{figure*}
\begin{center}
\subfloat[April/May 2013]{
     \includegraphics[scale=0.37]{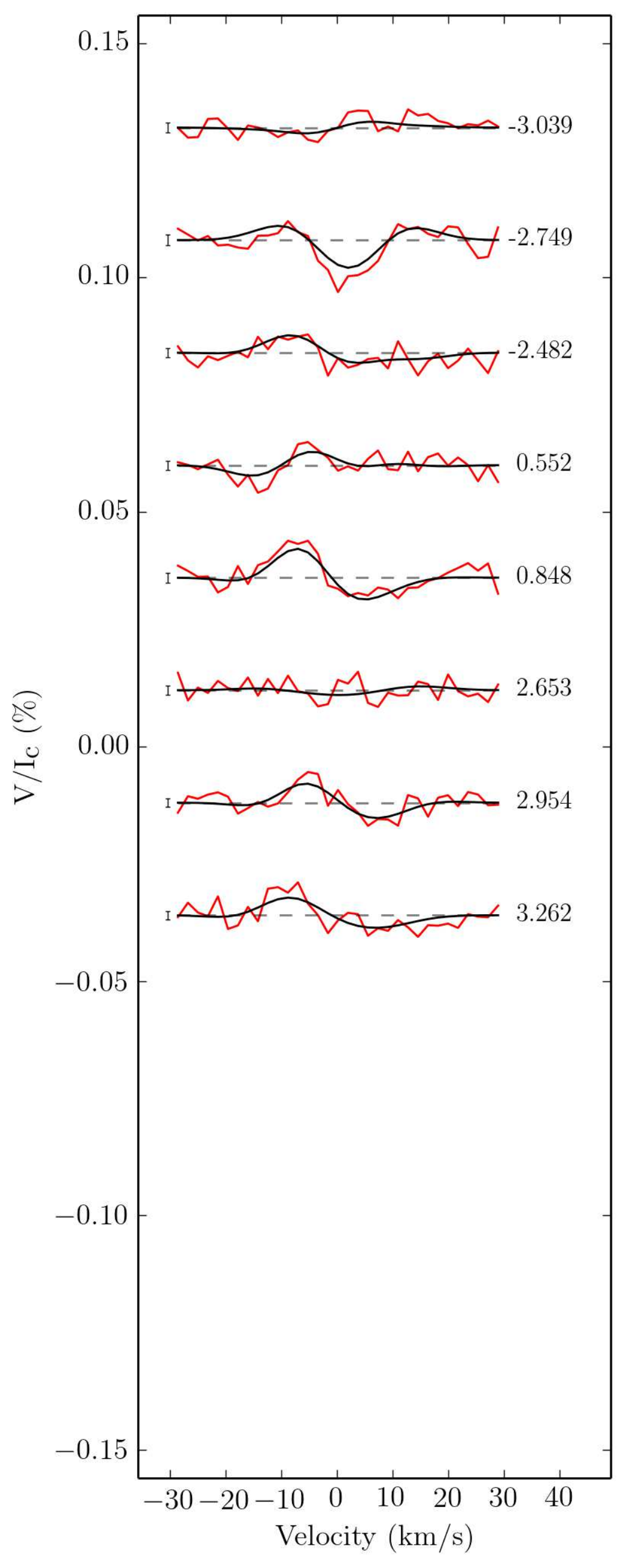}
}
\subfloat[December 2013]{
     \includegraphics[scale=0.37]{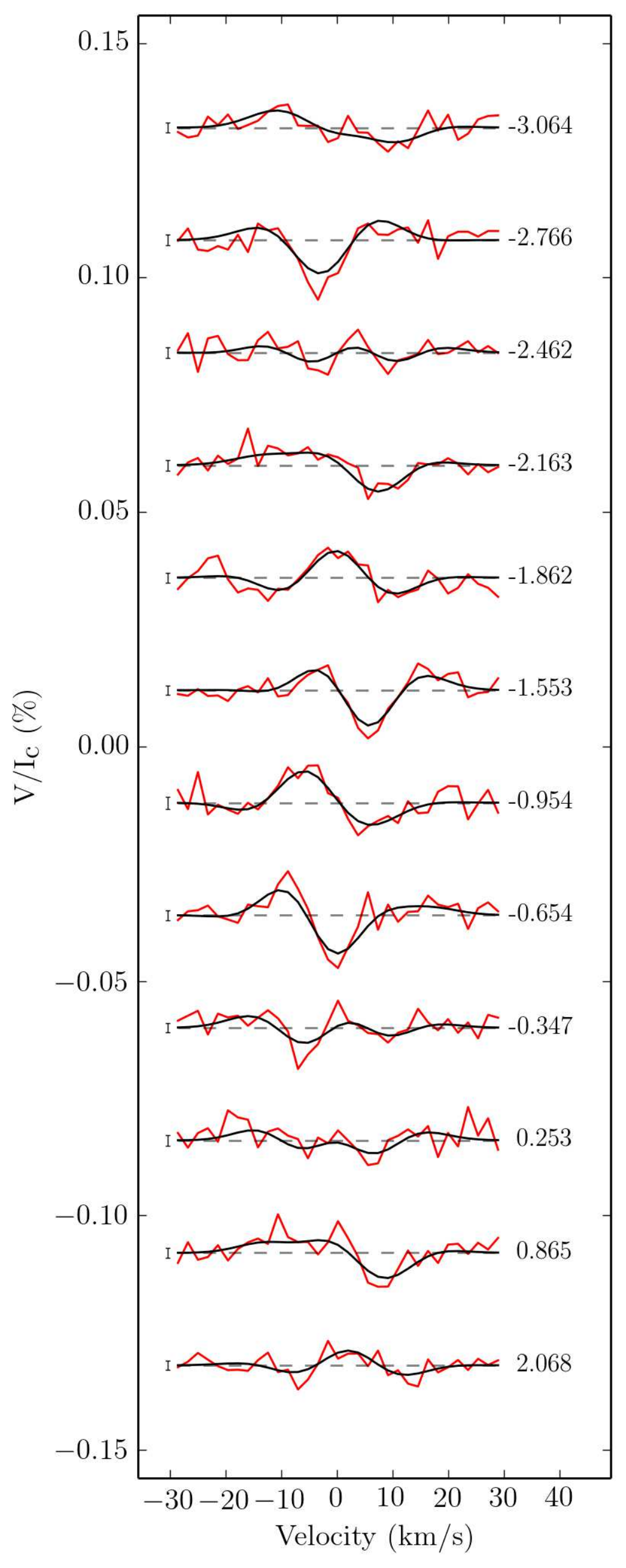}
} 
\subfloat[May 2014]{
     \includegraphics[scale=0.37]{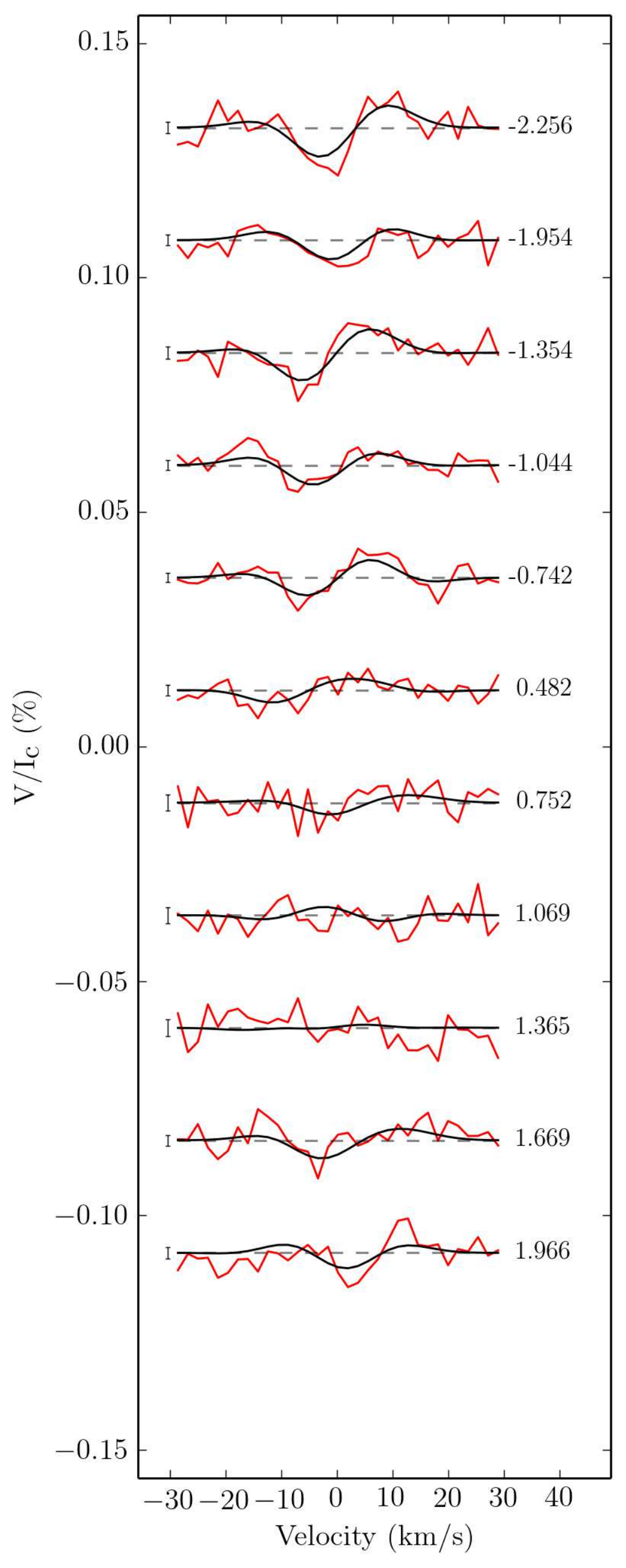}
}
\subfloat[January 2015]{
     \includegraphics[scale=0.37]{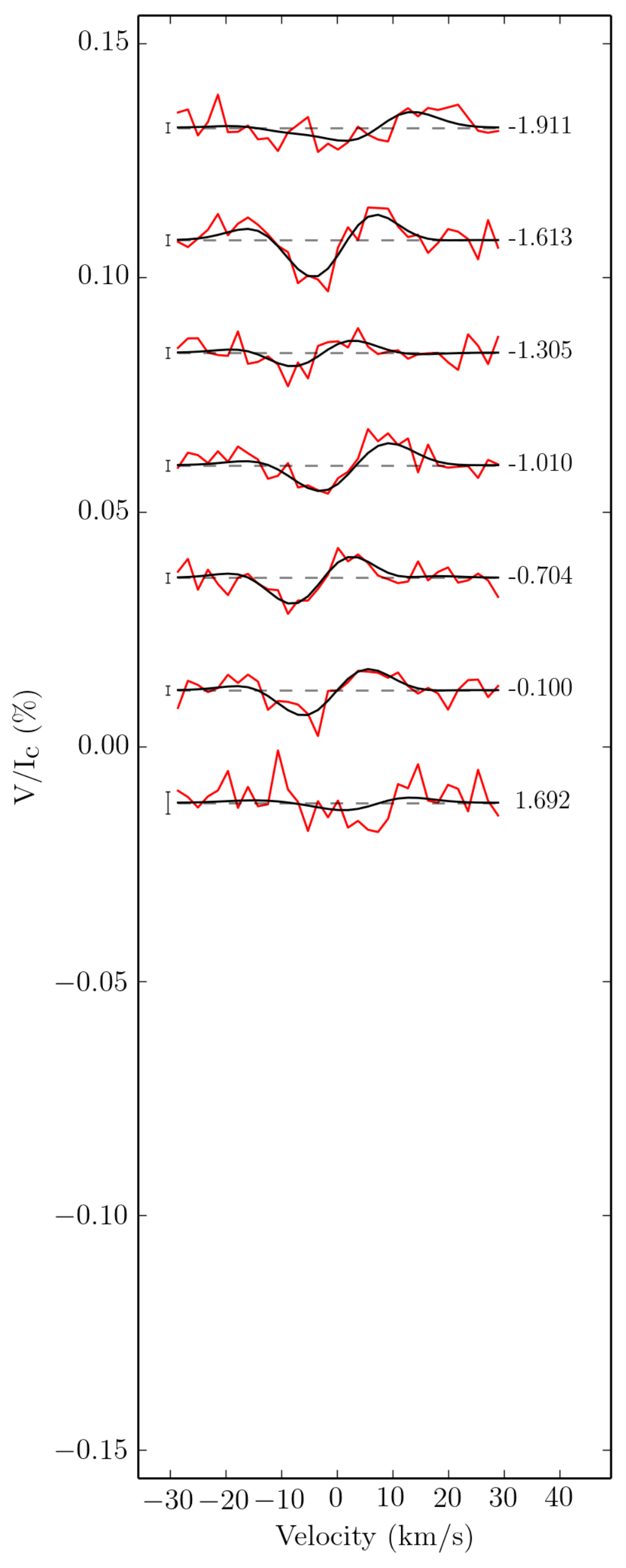}
}
\end{center}
  \caption{Circular polarization profiles of $\tau$ Boo for NARVAL observations, April 2013 through January 2015. The observed profiles are shown in red, while synthetic profiles are shown in black. On the left of each profile we show a $\pm1\sigma$ error bar. The rotational cycle of each observation is indicated on the right of each profile.}
  \label{fig:profiles}
\end{figure*}

\begin{figure*}
\subfloat{
     \includegraphics[ ]{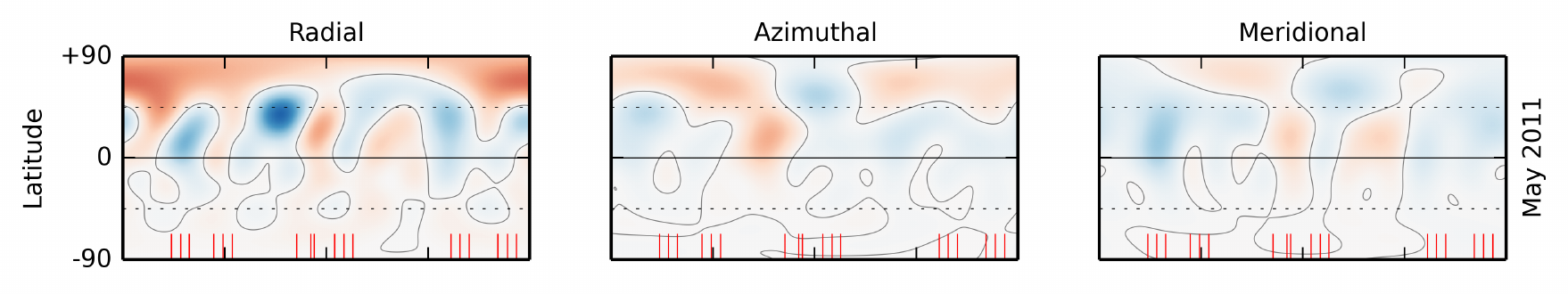}
}\vspace{-12pt}
\subfloat{
     \includegraphics[ ]{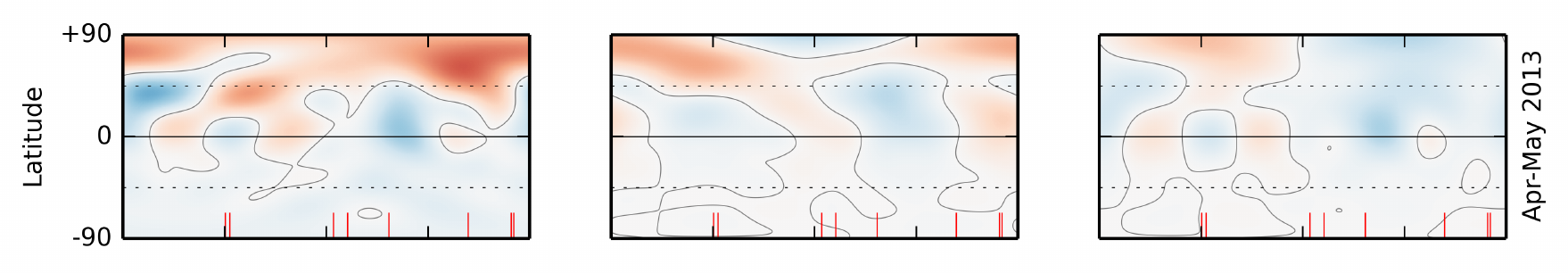}
}\vspace{-12pt}
\subfloat{
     \includegraphics[ ]{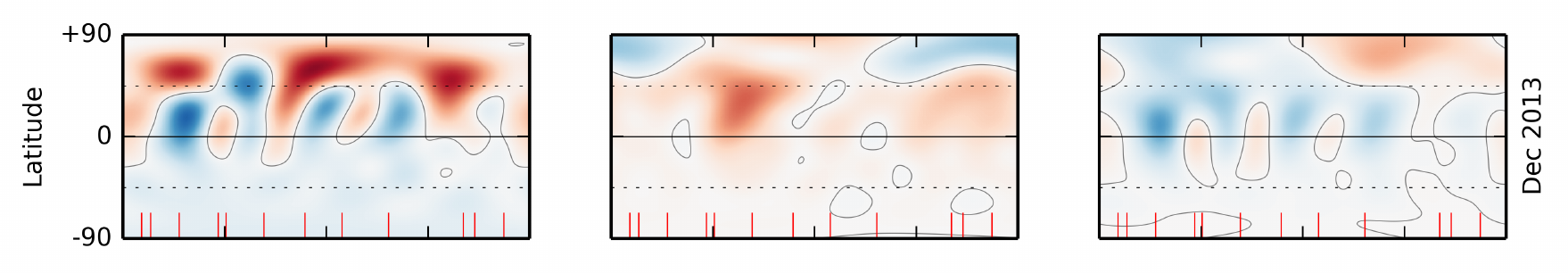}
}\vspace{-12pt}
\subfloat{
     \includegraphics[ ]{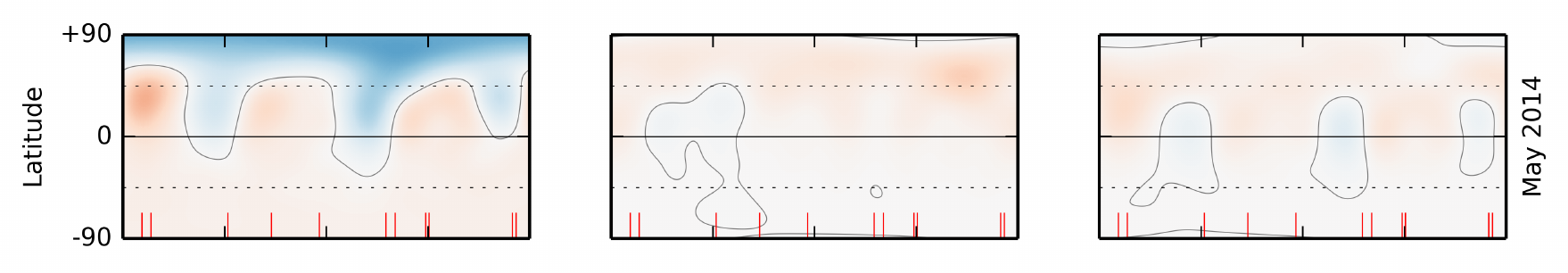}
}\vspace{-12pt}
\subfloat{
     \includegraphics[ ]{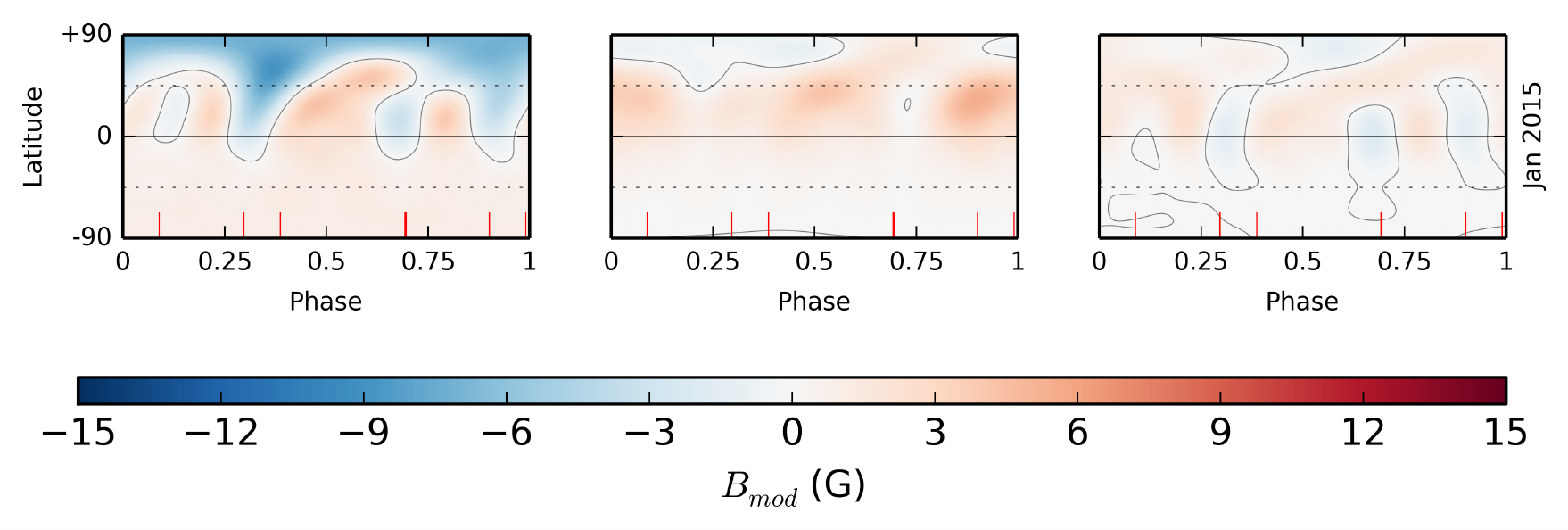}
}
\caption{Magnetic topology of $\tau$ Bo\"otis reconstructed from profiles in Fig.~\protect\ref{fig:profilesHARPS} and Fig.~\protect\ref{fig:profiles} (a) through (d). The radial, azimuthal and meridional components of the field (with magnetic field strength labelled in G) are depicted for May 2011 (top), April-May 2013 (second row), December 2013 (third row), May 2014 (fourth row), and January 2015 (bottom row).  The contour line indicates where $B_{mod}$ is zero.  The red ticks along the lower x-axes indicate the observational phases for each epoch.}
  \label{fig:maps}
\end{figure*}

\begin{figure*}
\begin{center}
\subfloat[02-Apr to 30-Apr]{
     \includegraphics[scale=0.35]{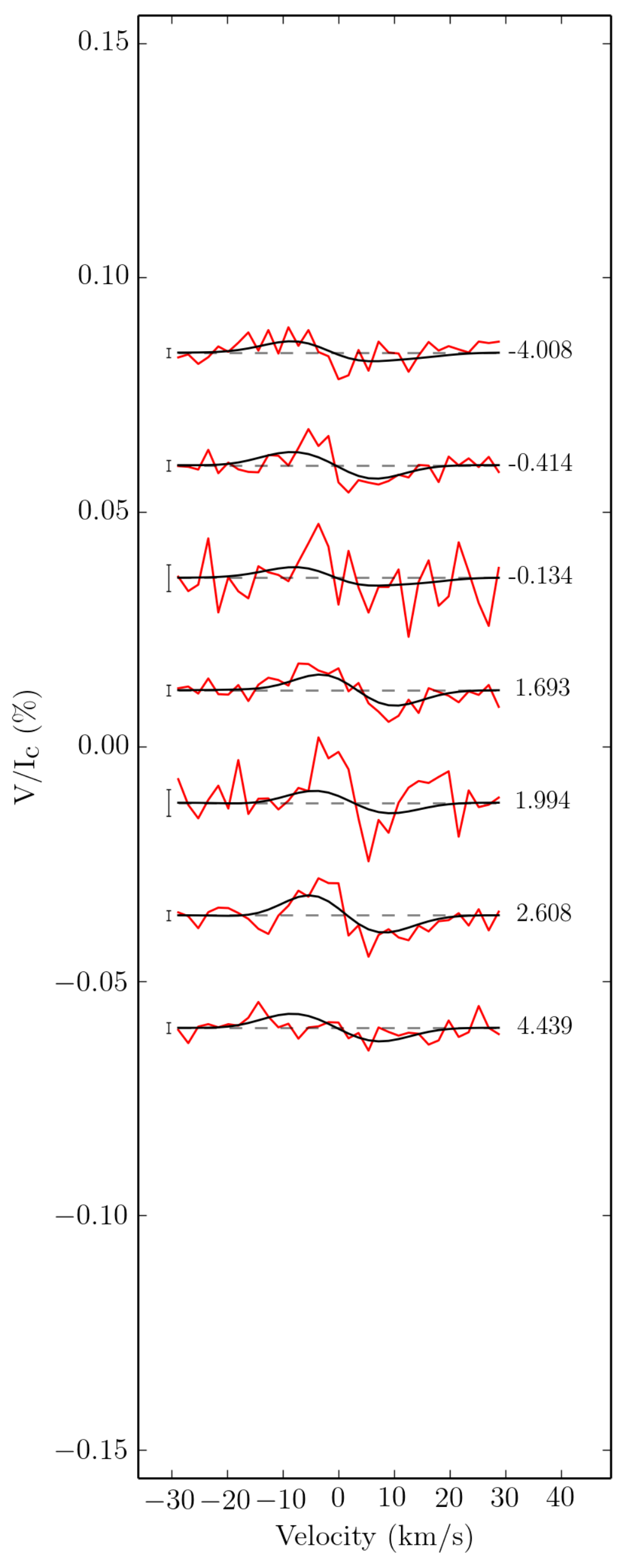}
}
\subfloat[13-Apr to 12-May]{
     \includegraphics[scale=0.35]{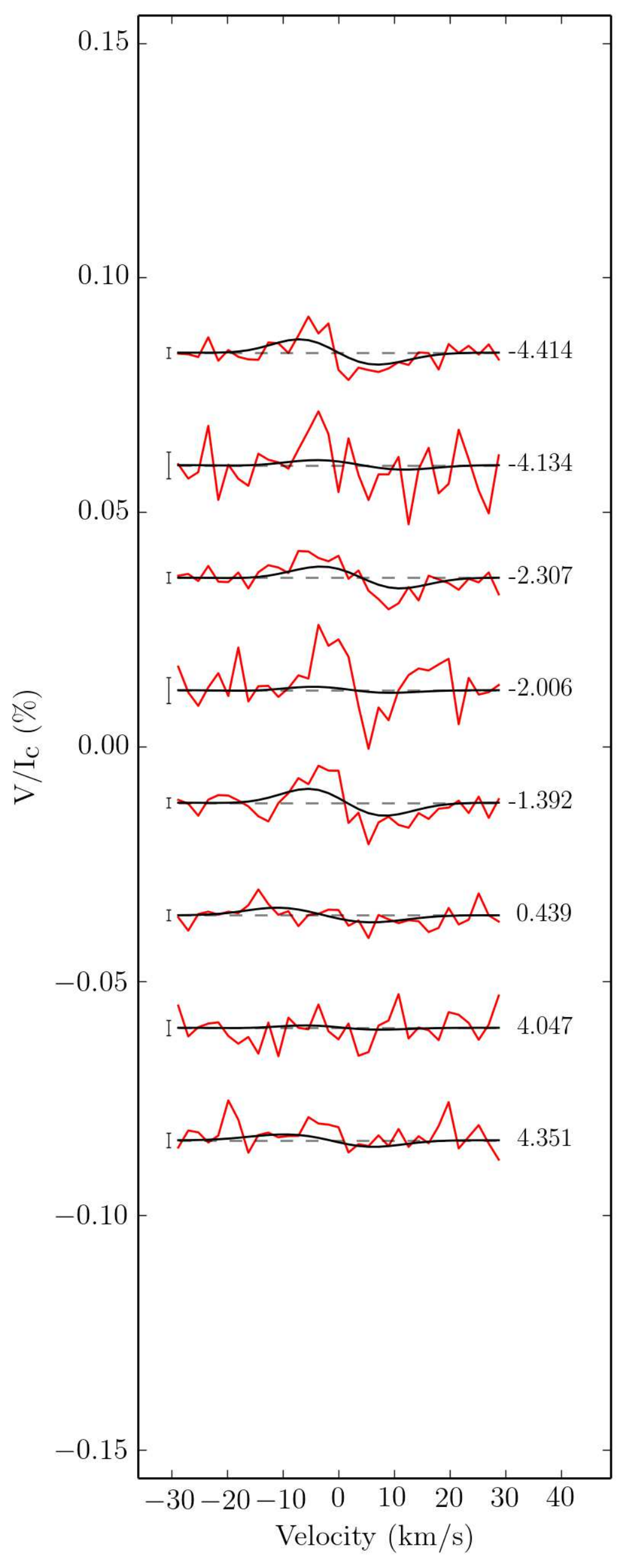}
}
\subfloat[20-Apr to 17-May]{
     \includegraphics[scale=0.35]{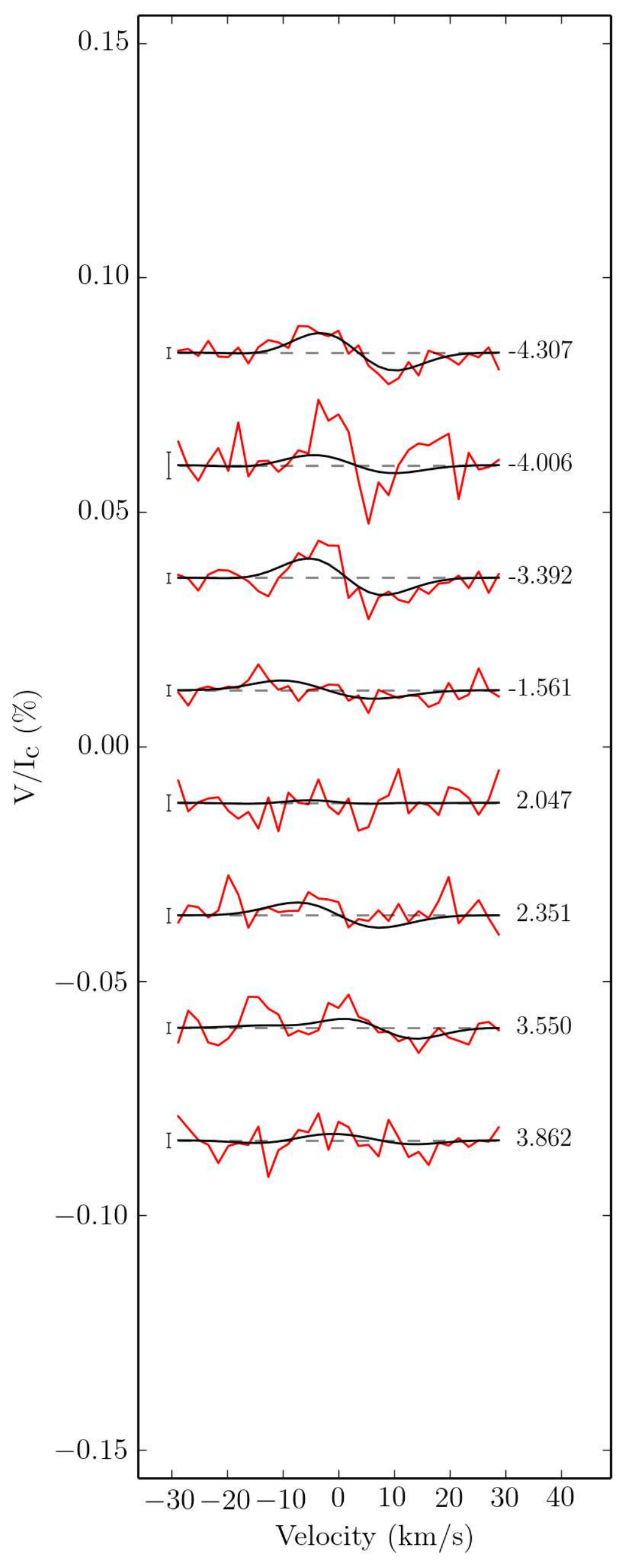}
} 
\subfloat[12-May to 27-May]{
     \includegraphics[scale=0.35]{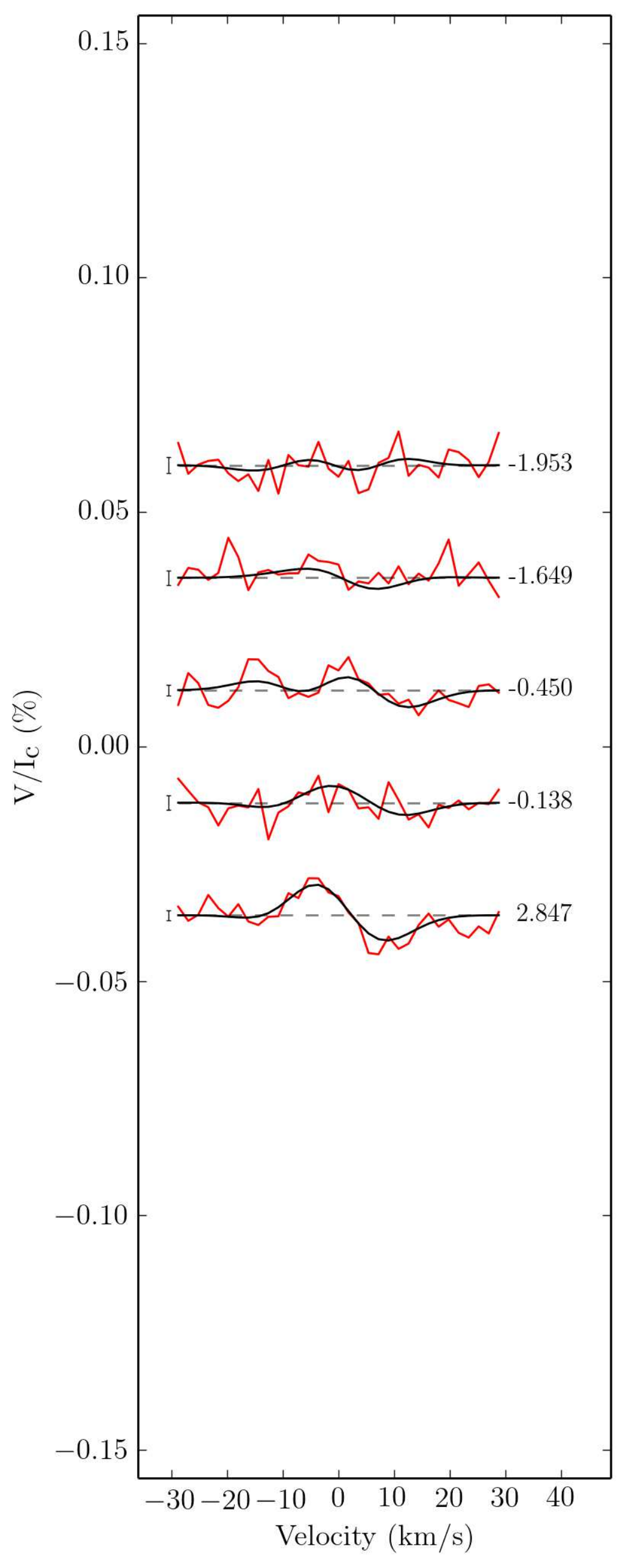}
}
\end{center}
  \caption{Circular polarization profiles of $\tau$ Boo from NARVAL April-May 2015.  As shown in Table~\protect\ref{table:marmaps}, four maps utilising a "sliding window" of approximately 8 rotations were used (apart from the final set of 5 rotations.  The sets overlap and phases are as per Table~\protect\ref{table:marmaps}. The observed profiles are shown in red, while synthetic profiles are shown in black. On the left of each profile we show a $\pm1\sigma$ error bar. The rotational cycle of each observation is indicated on the right of each profile.}
  \label{fig:profilesmar}
\end{figure*}

\begin{figure*}
\subfloat{
     \includegraphics[ ]{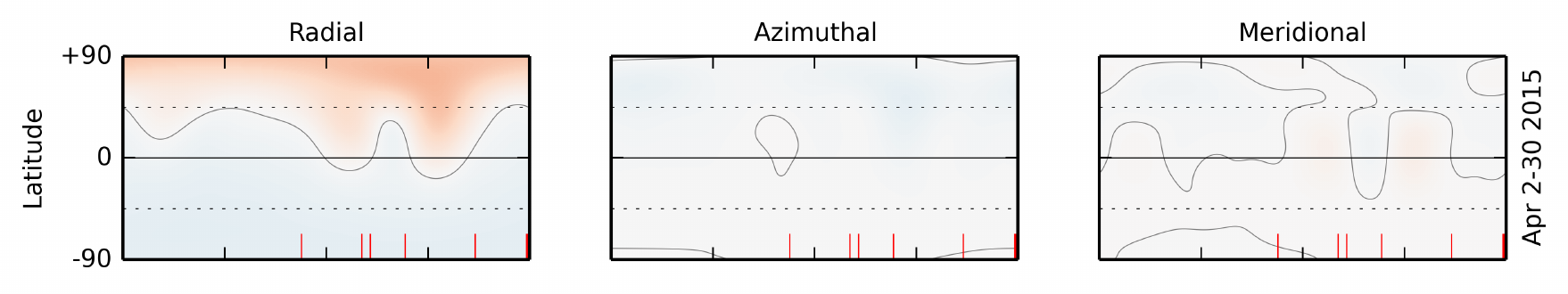}
}\vspace{-12pt}
\subfloat{
     \includegraphics[ ]{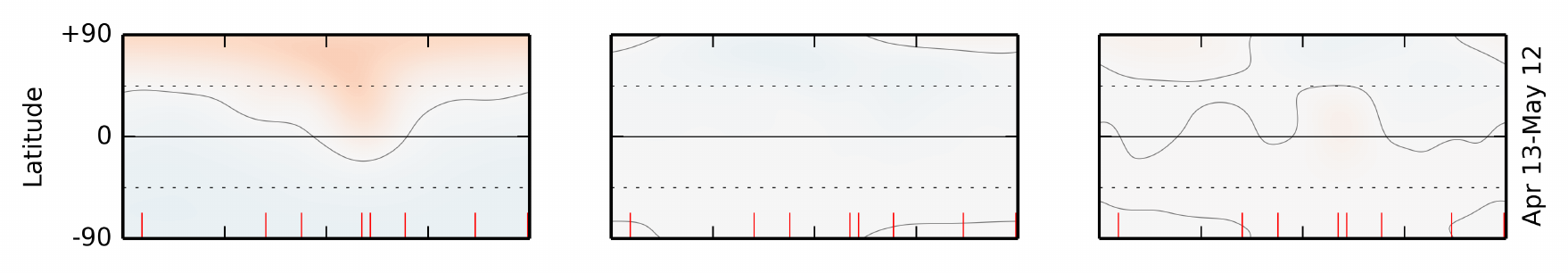}
}\vspace{-12pt}
\subfloat{
     \includegraphics[ ]{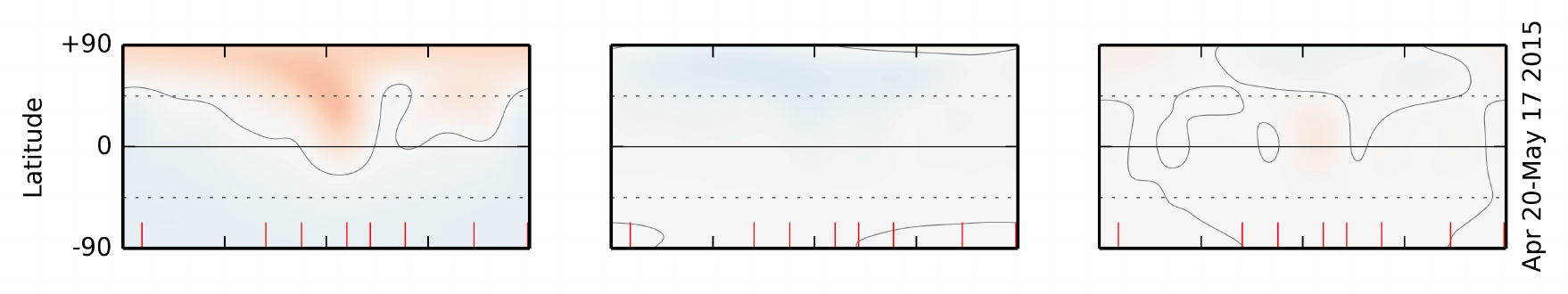}
}\vspace{-12pt}
\subfloat{
     \includegraphics[ ]{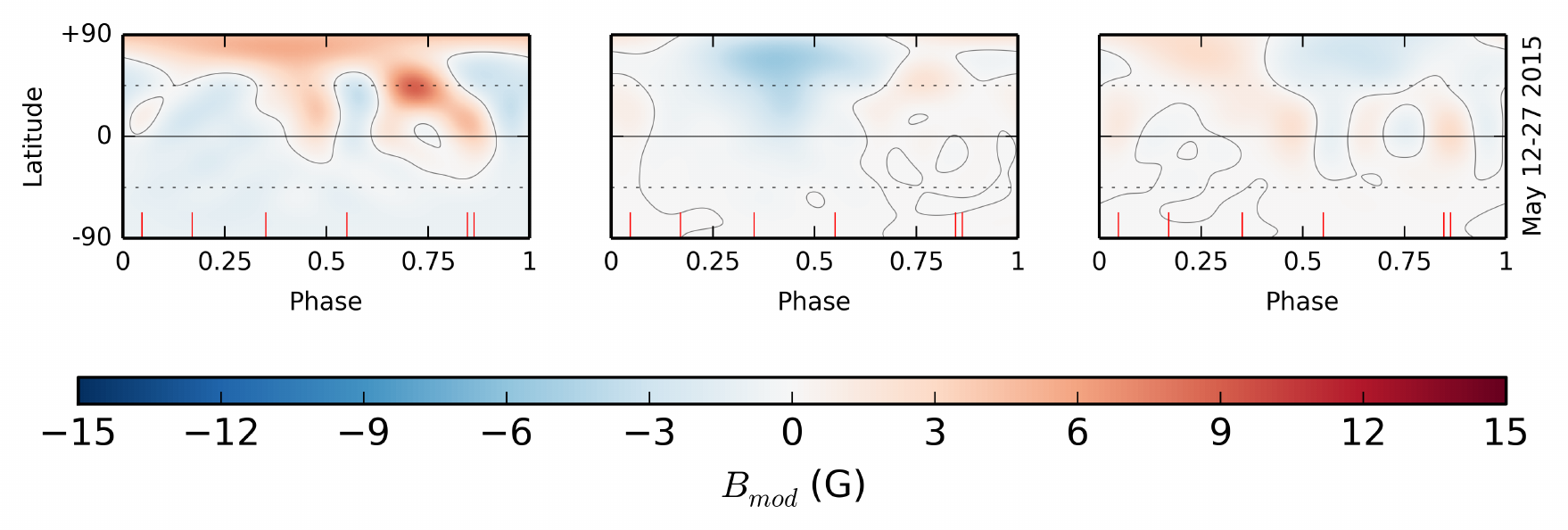}
}
\caption{Magnetic topology of $\tau$ Bo\"otis reconstructed from profiles in Fig.~\protect\ref{fig:profilesmar} (a) through (d). The radial, azimuthal and meridional components of the field (with magnetic field strength labelled in G) are depicted.  The contour line indicates where $B_{mod}$ is zero.  The red ticks along the lower x-axes indicate the observational phases for each epoch.}
  \label{fig:mapsmar15}
\end{figure*}

\subsubsection{Results - HARPSpol - May 2011}
\label{sec:map:harps}

Figure~\ref{fig:profilesHARPS} shows the observed and reconstructed profiles for May 2011.  Compared to the other epochs presented in this work, the magnetic signature is very small, and the observed profiles (in red) are relatively noisy, thus the reconstructed profiles begin fitting to the noise at a relatively high $\chi_{r}^{2}$.  While HARPSpol has a higher resolution than NARVAL (110000 cf. 65000), the spectral coverage is smaller and consequently there are significantly fewer spectral lines to use in LSD.  Nevertheless we observe, as per \citet{b4}, \citet{b5}, and \citet{b2}, that a weak magnetic field is manifest on the surface of $\tau$ Bo\"otis in the order of \SIrange{5}{10}{\gauss}.  The magnetic topology for May 2011 is shown in Fig.~\ref{fig:maps}, top row.  This topology was reconstructed with $\chi_{r}^{2} = 1.10$.

The magnetic field has evolved in configuration from the January 2011 observation presented in \citet{b5}.  The mean field is somewhat weaker at \SI{2.2}{\gauss} (although as noted before, a direct comparison of the HARPSpol B$_{mod}$ may be of limited utility given the differences in the $\chi_{r}^{2}$), and the percentage of the toroidal component of the field has remained around the same (from 18 to 20 percent).  In addition, the field configuration has become more complex, with 19 percent ($m=0$; value is $\sim25$ percent if using $m=\ell/2$) of the poloidal component in an axissymetric configuration compared to 37 percent, and only 26 percent of the poloidal component contained in modes $\ell \leq 2$ compared with 35 percent.  

\subsubsection{Results - NARVAL - April/May 2013 through January 2015}
\label{sec:map:narval1315}

Over a period from April 2013 until January 2015, four epochs of $\tau$ Bo\"otis were observed using NARVAL at TBL.  The reconstructed (black) and observed (red) magnetic profiles for these epochs are shown in Fig.~\ref{fig:profiles} (a), (b), (c) and (d).  In all four epochs, magnetic signatures are clearly observed, and the evolution of the signatures with rotation is clear.

All three sets of maps were reconstructed using the same inclination and $v\sin i$ parameters and $\chi_{r}^{2} = 0.95$.  The values of the differential rotation parameters ($\Omega_{eq}$, $d\Omega$) as derived in Section 4.2 for each epoch were used for each reconstruction.

It is clear from the maps (Fig.~\ref{fig:maps}, second, third, fourth and bottom rows) and from the parameters of the magnetic topology that the large-scale magnetic field on $\tau$ Bo\"otis evolved significantly over the eighteen month period.

Between April/May 2013 and December 2013, the field strength increased, and there was a negligible increase in the percentage of the calculated toroidal component.  The amount of poloidal axissymetry and the amount of field with modes of $\ell \leq 2$ decreased (Table~\ref{table:results}).  In short, the field increased slightly in complexity.

Between December 2013 and May 2014, a significant change occurred with a clear reversal of polarity in the radial field, coincident with a significant decrease in the percentage of the toroidal component (24\% to 13\%), an increase in the axissymetric poloidal component (30\% to 46\%) and a slight increase (40\% to 45\%) in the amount of the poloidal field with modes of $\ell \leq 2$.  After the reversal, the field became more poloidal and symmetric.

Between May 2014 and January 2015 as the next polarity reversal (expected between January and May 2015) approached, the toroidal component of the field once more increased (from 12\% to 30\%), the axissymetric component of the poloidal field dropped from 46 percent to 37 percent and the percentage of poloidal field in modes $\ell \leq 2$ decreased slightly from 45 to 41.  This represents another increase in the complexity of the field as the radial field reversal approaches.

There have now been three sets of sequential epochs of $\tau$ Bo\"otis observations which fall between radial field reversals.  The latest two epoch pairs (April/May and December 2013; May 2014 and January 2015) presented in this work show an increase in field strength and in complexity between the reversals based on the amount of poloidal axissymetric modes, modes with $\ell \leq 2$ and an increase in toroidal field.  The other inter-reversal epoch pair (May 2009/January 2010; \citet{b5}) of epochs showed an increase in field strength and a decrease in axissymetry of the poloidal field however the percentage of toroidal field did not increase in that case.

\subsubsection{Results - NARVAL - March though May 2015}
\label{sec:map:narval15}

Observations of $\tau$ Bo\"otis from March until May 2015 presented a challenge for mapping as the data was very sparse and spread over \SI{70}{\day} which corresponds to over 20 stellar rotations.  This presents a problem due to the potential for feature evolution.  In addition, only one or two observations per rotation leaves little information for determining differential rotation using the $\chi^{2}$-landscape technique.

The initial three observations (12-17 March 2015) were widely separated from the rest of the data set.  A crude analysis using these three observations confirmed that the expected polarity reversal had occurred, meaning that it had occurred between late January and early March.  The long time base of the March through May 2015 observations presented an opportunity to investigate this activity proxy minimum using the available data.

\begin{figure}
  \includegraphics[scale=0.45]{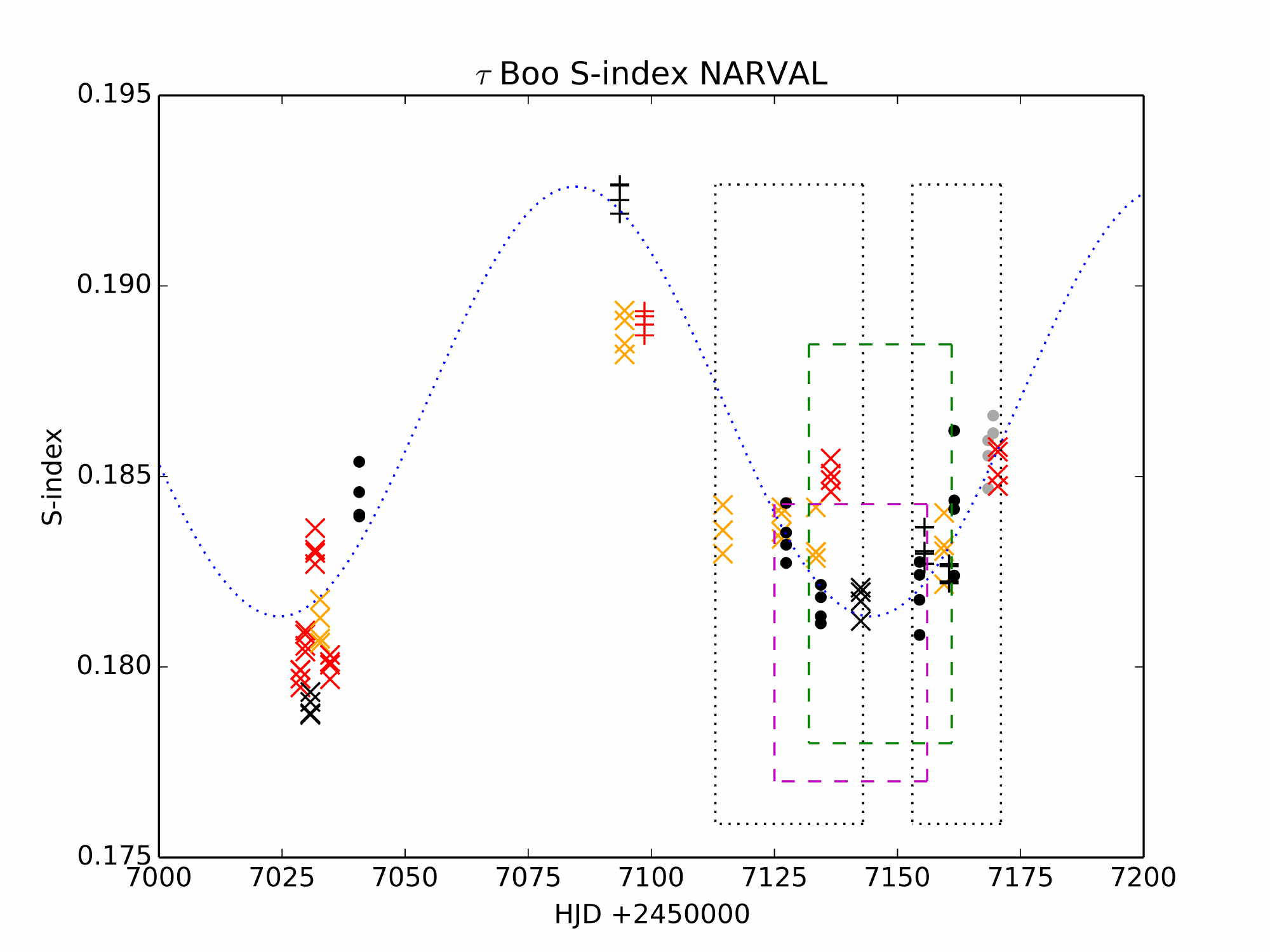}
  \caption{Ca II HK S-indices for $\tau$ Bo\"otis for January ($\sim 7030$~HJD+2450000) through May 2015.  NARVAL observations are shown as follows: red = definite detection; orange = marginal detection; black = no detection; grey = incomplete Stokes V sequence.  Marker shapes represent S/N near $\sim\SI{700}{\nano\meter}$ in the Stokes V spectrum: circles = S/N~$< 1000$; $+$ = $1000 <$~S/N~$<1300$; $\times$ = S/N~$>1300$.   A least-squares fit to the unweighted data (blue line) yields a period of $\sim\SI{120}{\day}$, similar to that reported in \protect\citet{b21}.  The leftmost black dotted rectangle indicates the observations used for the profiles in Figure~\protect\ref{fig:profilesmar}(a).  The magenta and green dashed rectangles correspond to  Figure~\protect\ref{fig:profilesmar}(b) and (c) respectively.  The rightmost dotted rectangle indicates the observations used for the profiles in Figure~\protect\ref{fig:profilesmar}(d). }
  \label{fig:marsdx}
\end{figure}

Utilising a sliding window of approximately 8 stellar rotations, the observations from April 2 through May 27 were split into four overlapping data sets (Fig.~\ref{fig:marsdx}; Table~\ref{table:marmaps}) and maps were generated.  Individual differential rotation measurements were made for each  of the four data sets and the measurements are summarised in Table~\ref{table:drmar}.  As can be seen, the $d\Omega$ values were near solid-body and poorly-defined for the first three data sets.  This is probably due to insufficient data for the technique~\citep{b33}.  For each map, $\chi_{r}^{2}$ of 0.95 was used.

The maps for the four data sets are shown in Figure~\ref{fig:mapsmar15} and the resulting field configuration information is presented in Table~\ref{table:resultsmar}.  While the sparseness of the data over this period and other observational biases such as poor S/N may contribute to the lack of features, if accurate, these maps show an intriguing progression through the activity proxy minimum.  The observable field strength decreases as the minimum in S-index is passed and increases dramatically as the S-index begins to rise.  In all cases, the strength is less than that from the January 2015 epoch prior to the dipole reversal despite being at a higher S-index.

Figure~\ref{fig:marsdx} does show there are several observations in this epoch that suffer from comparatively poor signal-to-noise (circles S/N~$< 1000$), where no magnetic signature is detected, which has an effect on the reconstruction.  In contrast, the January 2015 data (Fig.~\ref{fig:marsdx}; data located at $\sim$~HJD~2457030) where the measured S-indices are lower than the March-May data, apart from the final observation, the data is excellent with S/N above 1386.  Apart from the lowest S-index measurement, there are detections; mostly classified as definite.  Thus it may appear at first glance that the variation in the S-index may not correlate with the level of magnetic activity we detect with ZDI due to the effects of varying data quality.

\begin{figure}
  \includegraphics[scale=0.45]{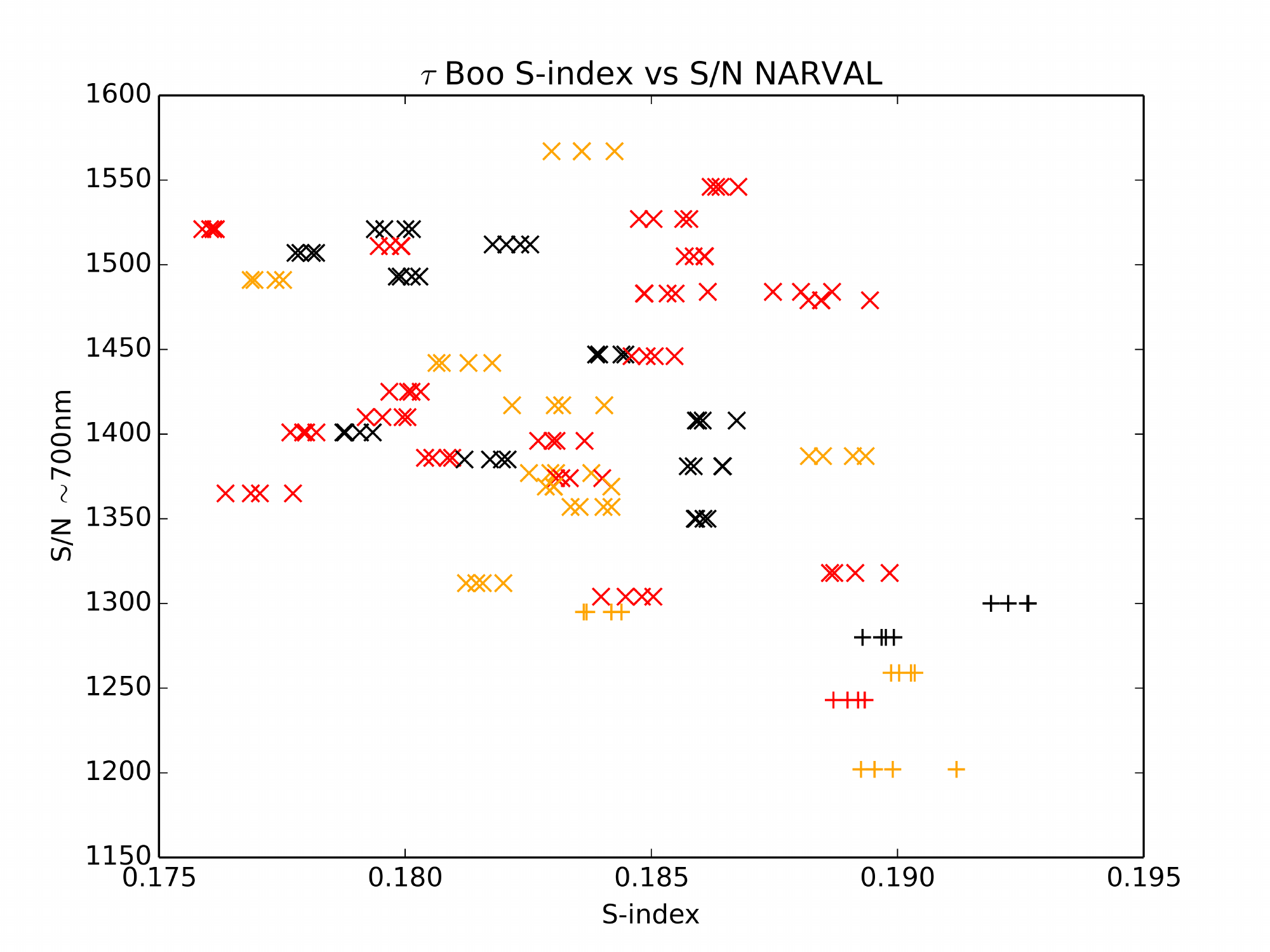}
  \caption{Signal to Noise measured at $\sim\SI{700}{\nano\meter}$ in the Stokes V spectra plotted against Ca \textsc{ii} HK S-indices for $\tau$ Bo\"otis for all NARVAL observations presented in this paper.  Observations are shown as follows: red = definite detection; orange = marginal detection; black = no detection.  Marker shapes represent S/N near $\sim\SI{700}{\nano\meter}$ in the Stokes V profile: $+$ = $1000 <$~S/N~$<1300$; $\times$ = S/N~$>1300$.  There is only one marginal detection below SN$\approx 1100$; all others are non-detections.  These observations excluded for clarity. }
  \label{fig:snsix}
\end{figure}

However, if we ignore the March-May 2015 observations with S/N$< 1000$, a pattern does appear where as the S-index decreases, the magnetic detections become marginal until finally there is a non-detection.  Further, examining all of the observations from the NARVAL epochs presented in this paper (Fig.~\ref{fig:snsix}, excluding S/N$< 1150$ where there is only a single marginal detection), it is clear that as the quality of the data improves, the proportion of marginal detections decreases, while non-detections only appear at lower S-indices.  Hence while the data quality is a factor, it is possible to say that chromospheric activity does exhibit a relationship to the observed magnetic activity of $\tau$ Bo\"otis.

This may only be within a particular chromospheric activity cycle.  Noting that although the S-indices for January 2015 were universally lower than March-May 2015, the star's magnetic field was stronger in January.  This may mean that there is a \textit{general} increase in magnetic activity as the star's magnetic cycle proceeds from one reversal to the next, while there is a smaller modulation following the chromospheric activity cycle of which there appear to be $\sim3$ per intra-reversal period ($\sim\SI{117}{\day}$:$\sim\SI{1}{\year}$).  However, the existing data is not detailed enough to make this statement definitively.  Indeed, examining previous epochs and given the uncertainty in exactly when the polarity reversal occurs, it is possible to speculate that the magnetic cycle may correspond to the chromospheric cycle.  Three reversals per year would potentially manifest as a yearly reversal due to observational cadences.  This is discussed further in Section~\ref{sec:conclusion}.

The field configuration for March-May 2015 appears to be generally highly symmetrical for the first three data sets, however there is very little energy in the spherical harmonics to work with during the S-index minimum. It is clear though that after the minimum, there is noticeably more toroidal field evident, and a significant decrease in poloidal axissymmetry and dipolar/quadrupolar modes.  This result further suggests that a magnetic activity cycle appears to be coincident with the chromospheric activity cycle.

Given that $\phi_{rot} = 0$ is synchronised in each map, it is possible to see features migrating across the star between March and May 2015, probably due to differential rotation. In addition, in the final two data sets, the emergence of the prominent feature at $+\ang{45}$ near $\phi_{rot} = 0.75$ with time is clear.  While this may be due to observational effects, it also may be due to increased intensity leading to feature evolution.  Notably, maps 1, 3 and 4 all exhibit active regions near $\phi_{rot} = 0.75$.  Coincidentally, this is near the active longitude suggested by \citet{b16}.

\begin{table}
 \centering
  \caption{Use of exposures from April and May 2015 (Table~\protect\ref{table:obsNARVAL2015}) and their use in each map reconstruction.  The rotational cycle used for calculating 0.0 (from Equation~\protect\ref{eq:ephem}) is shown for each map. ($^{*}$ - Observation on 19-May with very poor S/N was not used in the reconstruction)}
\label{table:marmaps}
  \begin{tabular}{@{}cccrrrr@{}}
  \hline

&  & Map 1 & Map 2 & Map 3 & Map 4 \\
Obs. & Cycle & $\phi_{rot}$ & $\phi_{rot}$ & $\phi_{rot}$ & $\phi_{rot}$ \\
\hline
 &  &  &  &  &  \\
02-Apr-2015 & 1105.992 & -4.008 &  &  &  \\
13-Apr-2015 & 1109.586 & -0.414 & -4.414 &  &  \\
14-Apr-2015 & 1109.866 & -0.134 & -4.134 &  &  \\
20-Apr-2015 & 1111.693 & 1.693 & -2.307 & -4.307 &  \\
21-Apr-2015 & 1111.994 & 1.994 & -2.006 & -4.006 &  \\
23-Apr-2015 & 1112.608 & 2.608 & -1.392 & -3.392 &  \\
30-Apr-2015 & 1114.439 & 4.439 & 0.439 & -1.561 &  \\
11-May-2015 & 1118.047 &  & 4.047 & 2.047 & -1.953 \\
12-May-2015 & 1118.351 &  & 4.351 & 2.351 & -1.649 \\
16-May-2015 & 1119.550 &  &  & 3.550 & -0.450 \\
17-May-2015 & 1119.862 &  &  & 3.862 & -0.138 \\
19-May-2015 & 1120.170 &  &  &  & $^{*}$ \\
27-May-2015 & 1122.847 &  &  &  & 2.847 \\
 &  &  &  &  &  \\
$\phi_{rot}=0$ Cycle &  & 1110 & 1114 & 1116 & 1120 \\
\hline
\end{tabular}
\end{table}

\begin{table}
 \centering
  \caption{Summary of measured differential rotation parameters for $\tau$ Bo\"otis for the four maps March through May 2015.  Note the anti-solar $d\Omega$ for the first three epochs, due to poor coverage and/or low detectability of the magnetic field.}
\label{table:drmar}
  \begin{tabular}{@{}lcc@{}}
  \hline
   Epoch & $\Omega_{eq}$ & $d\Omega$  \\
        & \SI{}{\radian\per\day}  & \SI{}{\radian\per\day} \\
 \hline \\[-1.5ex]
2015 02 Apr - 30 Apr & $2.01^{+0.01}_{-0.15}$ & $-0.07^{+0.07}_{-0.04}$ \\[3pt]
2015 13 Apr - 12 May & $1.88^{+0.15}_{-0.04}$ & $-0.12^{+0.13}_{-0.09}$ \\[3pt]
2015 20 Apr - 17 May & $1.93^{+0.01}_{-0.01}$ & $-0.04^{+0.04}_{-0.03}$ \\[3pt]
2015 12 May - 27 May & $1.92^{+0.02}_{-0.06}$ & $+0.17^{+0.16}_{-0.14}$ \\[3pt]
\hline
\end{tabular}
\end{table}

\section[]{Discussion and Conclusions}
\label{sec:conclusion}

The March through May 2015 observational epoch makes clear that $\tau$ Bo\"otis is near the limit of our ability to apply ZDI.  A small magnetic survey of $\sim20$ planet hosting stars utilising NARVAL (Mengel et. al., in preparation) and the wider BCool survey \citep{b18} finds that fainter stars (these surveys are to $V\sim 9$) with an S-index less than $\sim0.2$ usually have magnetic fields too weak to detect.  $\tau$ Bo\"otis is thus unusual in this regard, due to its brightness ($V\approx4.5$) compared to the wider sample allowing for adequate signal-to-noise for ZDI.  When $\tau$ Bo\"otis is at the nadir of its chromospheric activity cycle, the magnetic field is on the very limit of where NARVAL can with confidence detect the Zeeman signature. 

Despite this limitation. the evolution of the large-scale magnetic field of $\tau$ Bo\"otis is apparent in the results we present here over two timescales and (what appear to be) two cyclic periods.  Most significantly, $\tau$ Bo\"otis appears to undergo polarity switches on a regular basis. \citet{b5} notes that the polarity switch appears to be a phenomenon occurring every \SI{360}{\day}, and these latest observations would seem to confirm this hypothesis.  The regular reversals appear to occur between January and March in each calendar year.  Additionally, magnetic energy appears to rise and fall in a shorter cycle coincident with the S-index of the star.

\begin{figure}
\begin{center}
\includegraphics[scale=0.45]{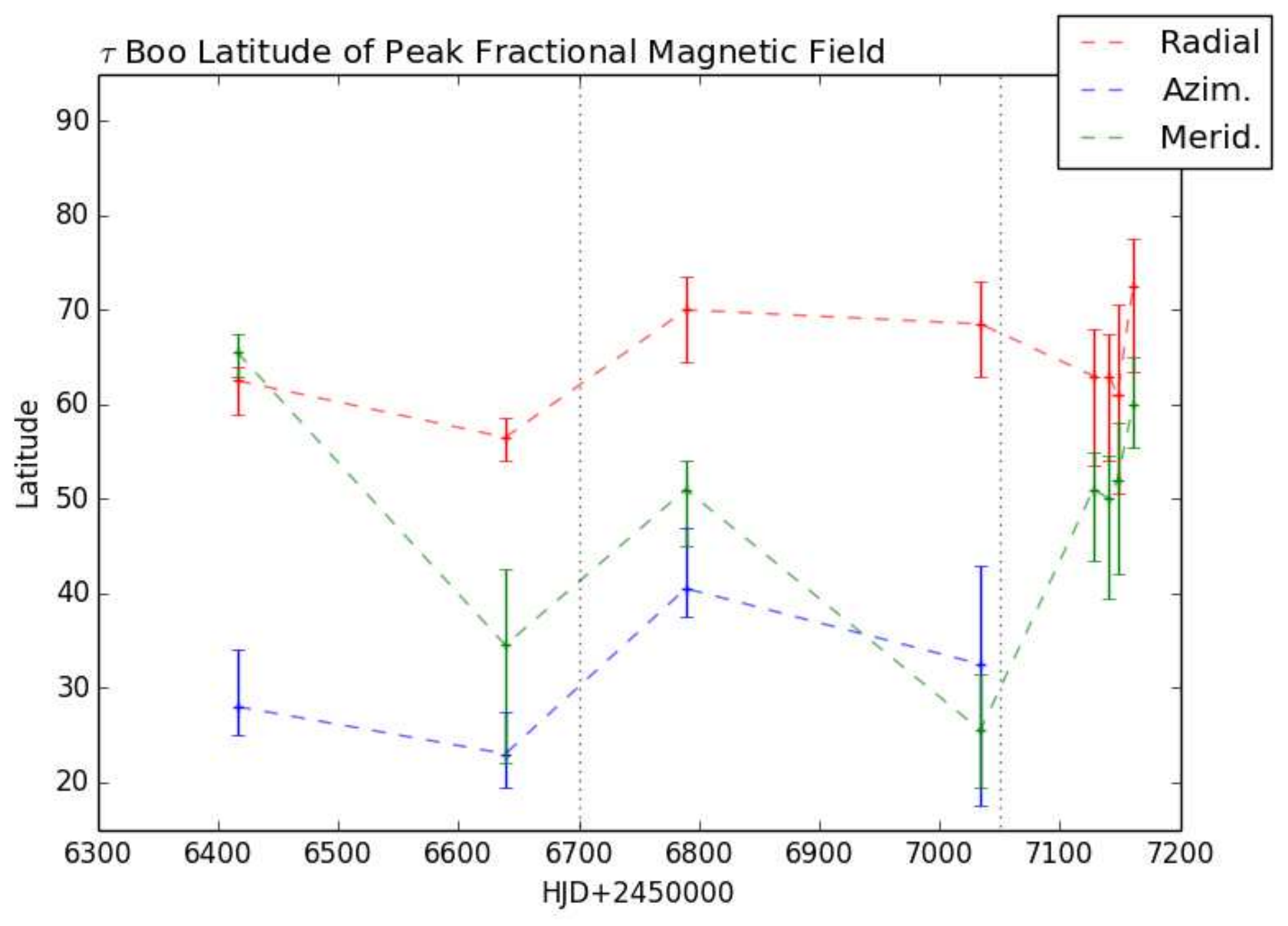}
\end{center}
\caption{Latitude of peak fractional magnetic field components for epochs April-May 2013 through January 2015 with variation bars from Figure~\ref{fig:fracmag} and for the March through May 2015 epoch.  The black dotted lines indicate a radial field reversal has taken place.  Azimuthal field is not included for the March though May 2015 epoch for clarity as the small amount of azimuthal field is spread across a wide area, making the variation bars extremely large.}
\label{fig:fracmaglat}
\end{figure}

The evolution of the magnetic field during the period between reversals is illuminating insofar as the behaviour of the large scale field broadly follows what one would expect in a star such as the Sun.  In Fig.~\ref{fig:fracmag}, the fractional magnetic field for each of the three components at each latitude is shown for these three epochs.  Between April/May 2013 and December 2013, the intensity of the radial and azimuthal components increases and the latitude at which the peak intensity is observed decreases towards the equator (Fig~\ref{fig:fracmag} (a) and (b)). After the polarity switch, the radial and azimuthal fields have a relatively lower intensity and their peak intensity is at latitudes close to the pole (Fig.~\ref{fig:fracmag} (c)).  This cycle then repeats between the two observed radial field reversals (Fig.~\ref{fig:fracmaglat}). This pattern of intensifying azimuthal field at lower latitudes approaching a reversal was observed between May 2009 and January 2010, the only similar set of observations taken between a pair of reversals.  It should be noted that with poorer phase coverage, the potential  uncertainty in the latitudes of recovered features increases, particularly in the azimuthal and and meridional components.  This is because with sparser phase coverage, we are less likely to observe either the exact points of entry or exit (or indeed both) of features on the visible stellar surface.  The level of this uncertainty is difficult to characterise, thus further observations with higher cadence and denser phase coverage may be required to absolutely confirm the nature of the latitudinal migration of features we propose here. 

As a polarity shift approaches, the field also becomes less axissymetric in the poloidal component and more complex, with slightly less of the poloidal field in modes corresponding to dipolar and quadrupolar components.  After the polarity shift occurs, the amount of the toroidal field decreases significantly and the field becomes simpler; more axissymetric and more strongly dipolar/quadrupolar.  (Note that between May 2009 and January 2010 \citep{b5}, the percentage of axissymetric modes similarly dropped precipitously, however, the complexity of the field decreased slightly differing from the 2013 epochs).

The $\sim\SI{117}{\day}$ period of the Ca \textsc{ii} H\&K index coupled with a posited $\sim$yearly radial field polarity reversal suggests a 3:1 periodic relationship between the cycles.  Reversals appear to occur at or near the peak of every third chromospheric cycle.  While the sequence of maps by \citet{b4,b5,b2,b3} indicates that the period of reversal appear to be yearly, this may not necessarily the case.  On the face of it, $\tau$ Bo\"otis has an approximately \SI{720}{\day} magnetic cycle, which is mostly, but not completely, an analogue of a solar-like cycle.  Our results from March through May 2015 show that the timing of the observational epochs for $\tau$ Bo\"otis may in fact disguise a much faster cycle, such as the \SI{240}{\day} period found by \citet{b5}.

\begin{figure}
\begin{center}
\includegraphics[scale=0.45]{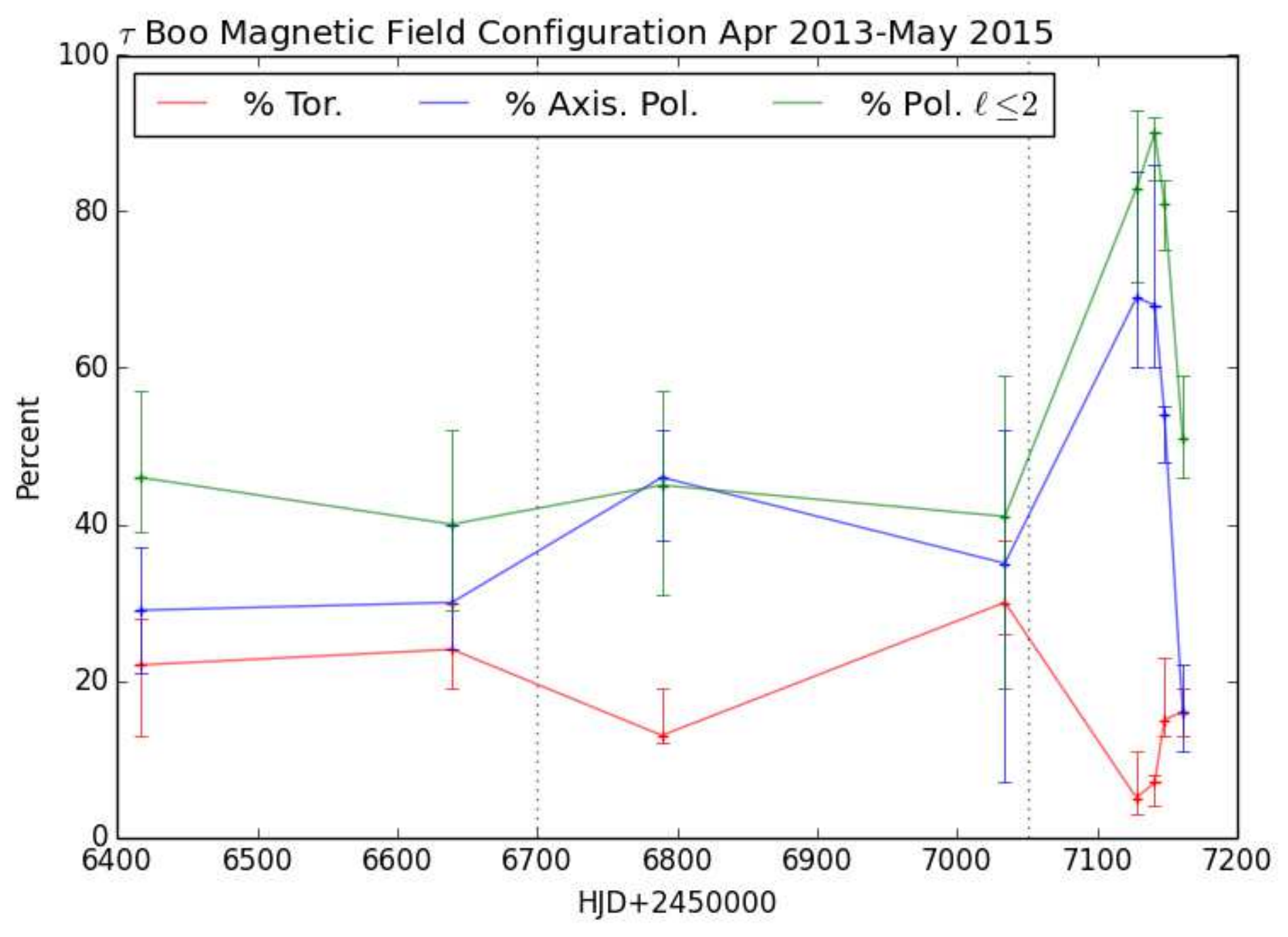}
\end{center}
\caption{Plot of magnetic field topology from Tables~\protect\ref{table:results} and \protect\ref{table:resultsmar}. The black dotted lines indicate approximately where a radial field reversal has taken place.  The rapid evolution of the measured field components after HJD~2457100 (Mar-May 2015) is coincident with the chromospheric cycle and suggests the magnetic cycle may be more rapid than the $\sim\SI{720}{\day}$ cycle previously assumed.}
\label{fig:magconfig}
\end{figure}

\begin{figure}
\begin{center}
\includegraphics[scale=0.45]{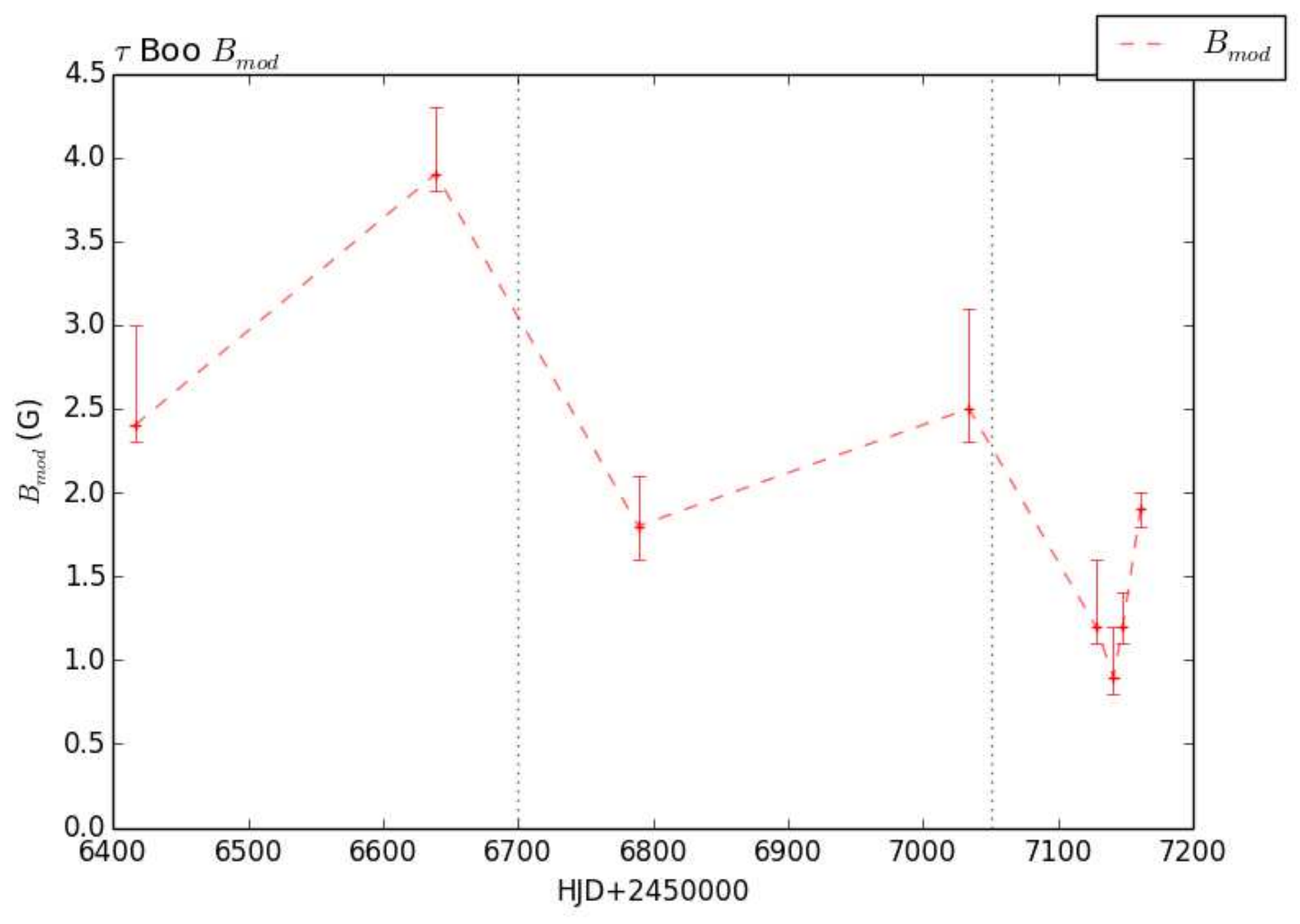}
\end{center}
\caption{Plot of mean magnetic field strength ($B_{mod}$) from Tables~\protect\ref{table:results} and \protect\ref{table:resultsmar}. The black dotted lines indicate approximately where a radial field reversal has taken place.}
\label{fig:magbmod}
\end{figure}

\begin{figure*}
\begin{center}
\subfloat[April/May 2013]{
     \includegraphics[scale=0.4]{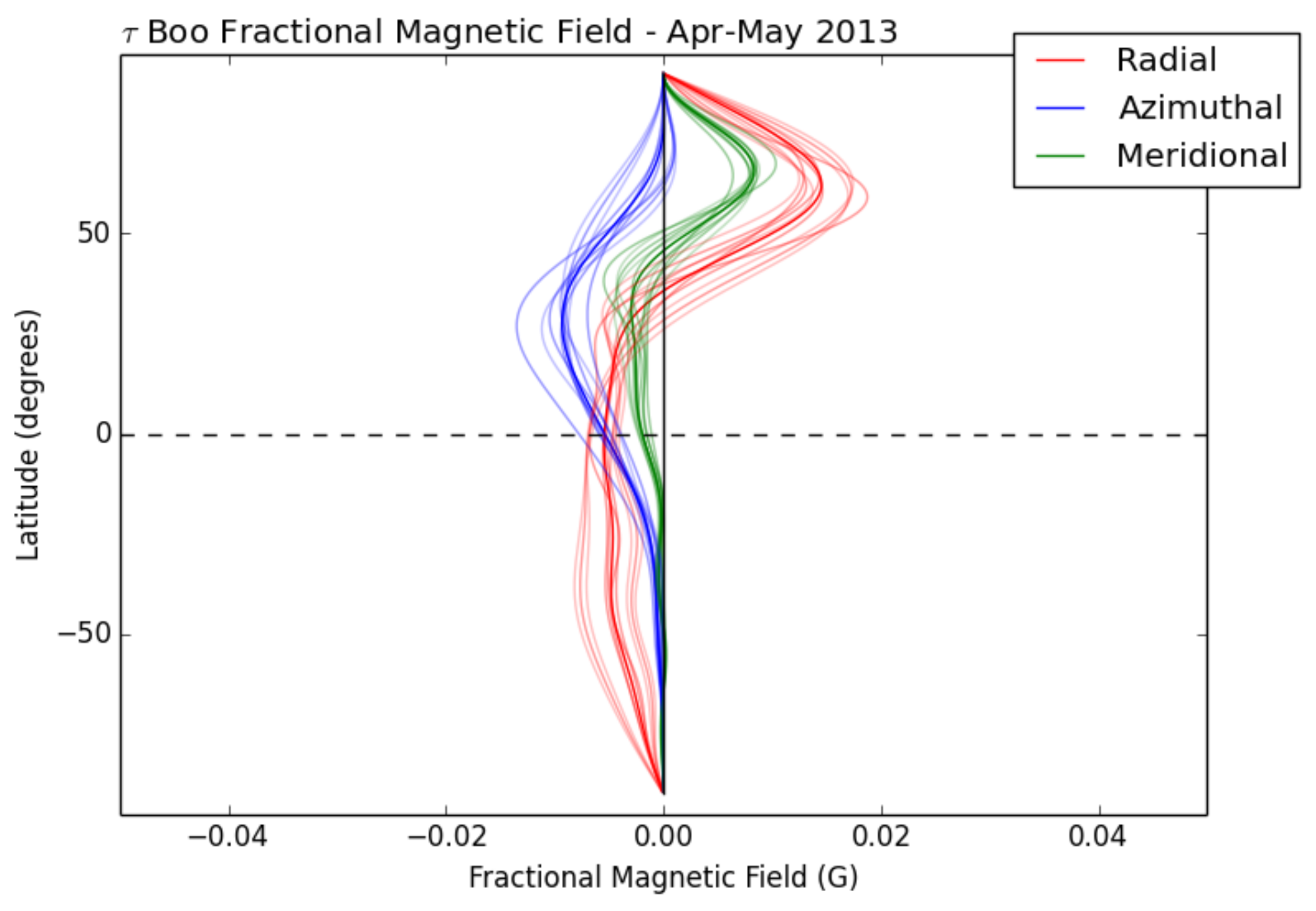}
}
\subfloat[December 2013]{
     \includegraphics[scale=0.4]{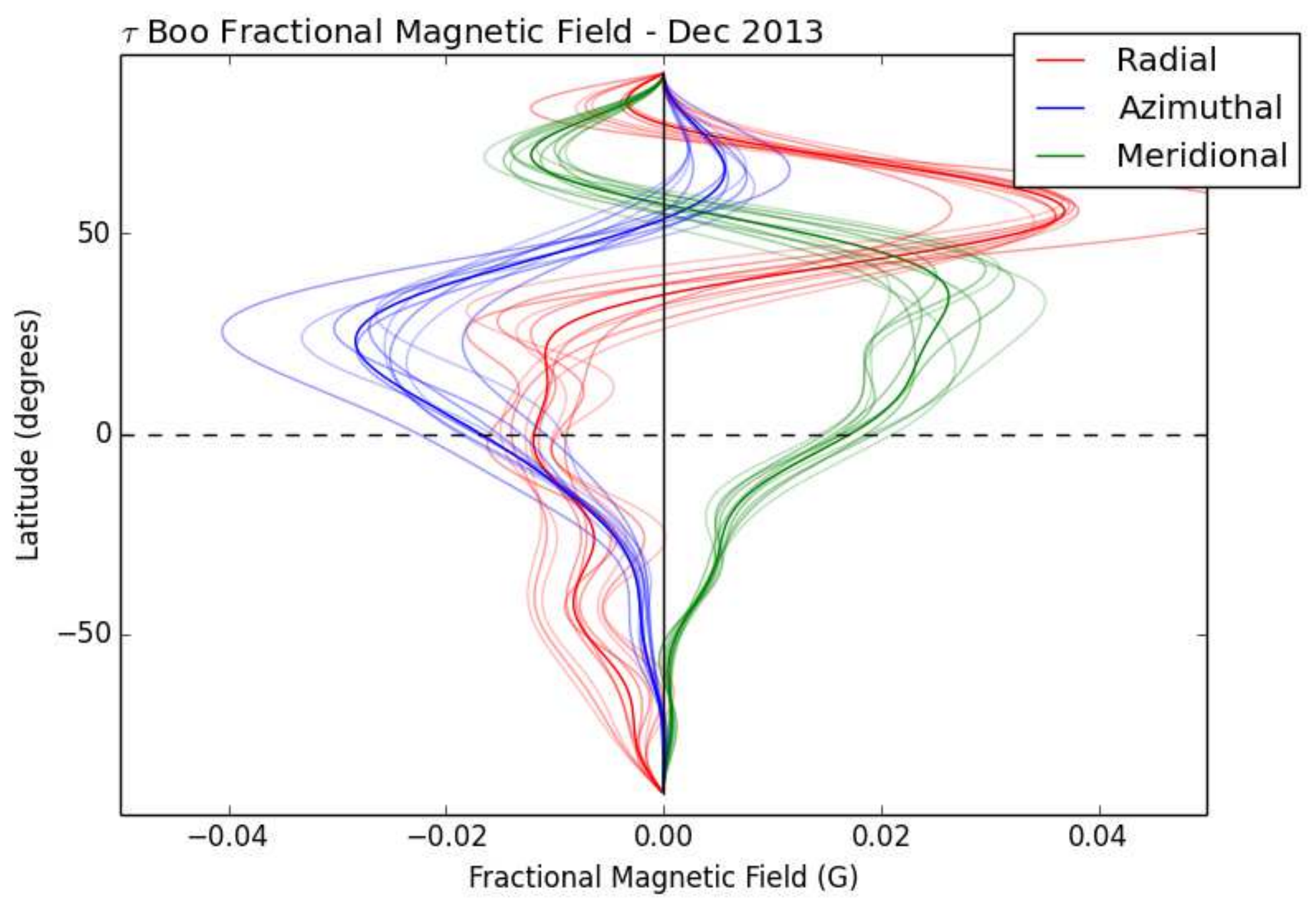}
} \\
\subfloat[May 2014]{
     \includegraphics[scale=0.4]{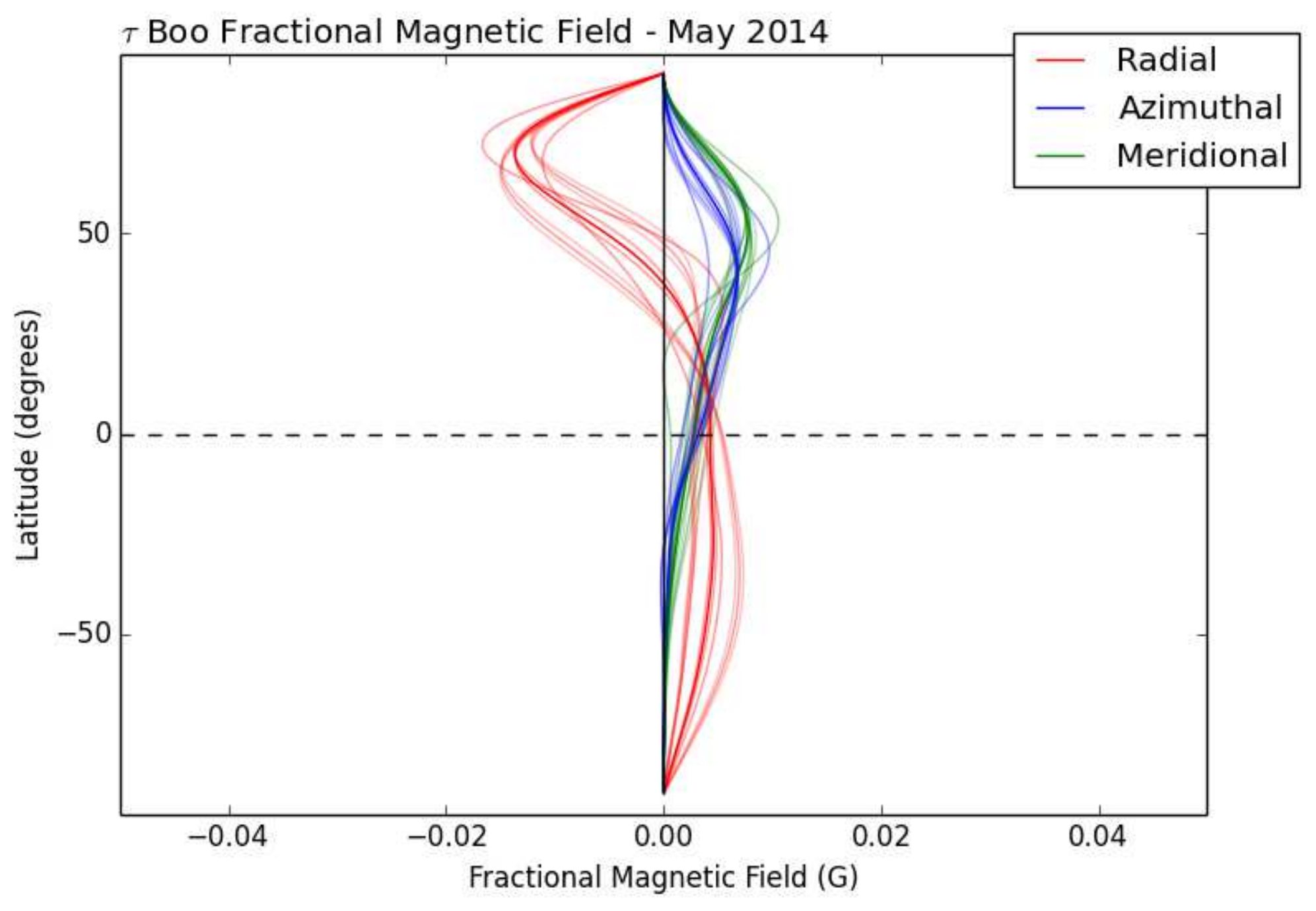}
}
\subfloat[January 2015]{
     \includegraphics[scale=0.4]{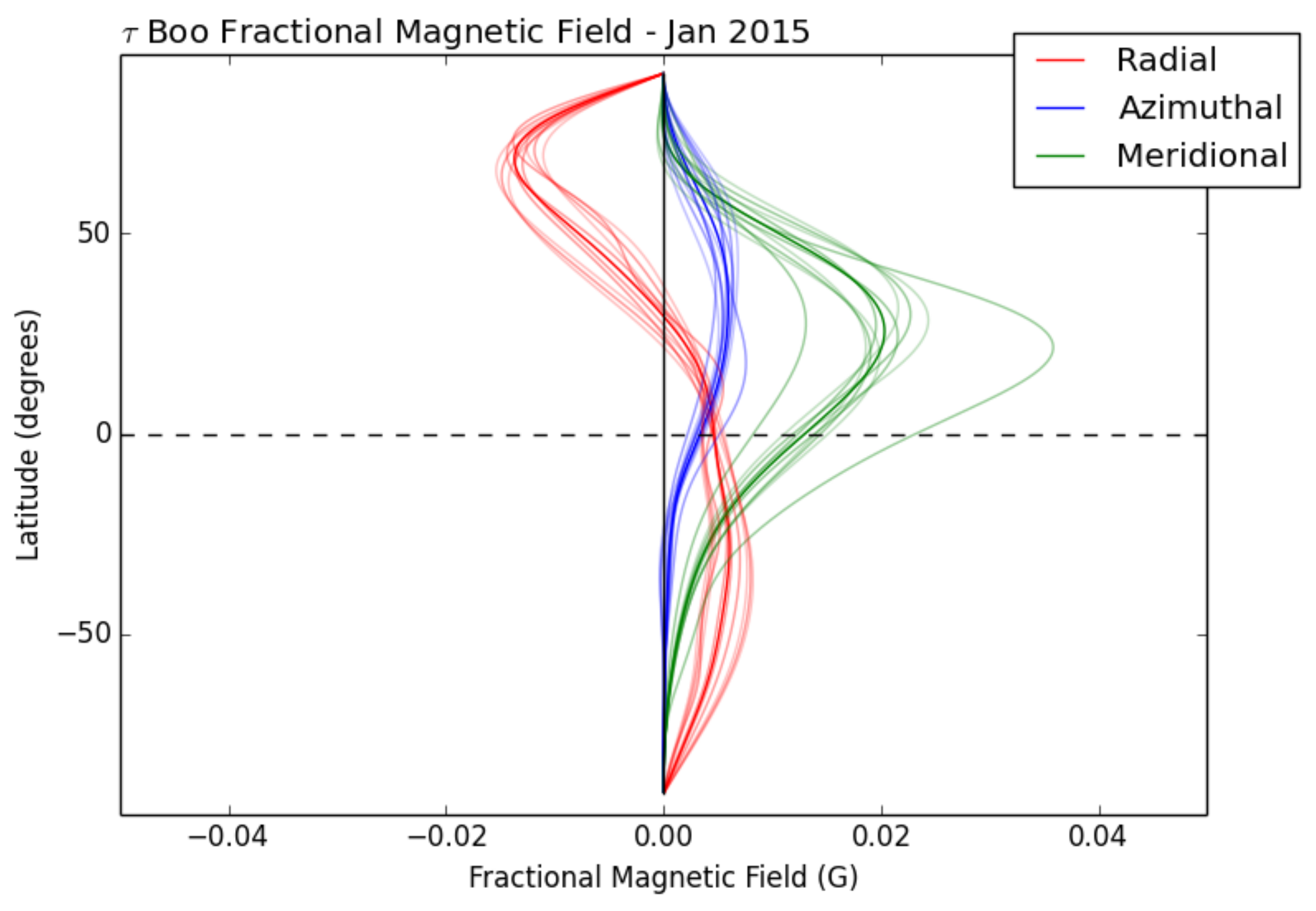}
}
\end{center}
  \caption{Fractional magnetic field by latitude for (a) April/May 2013, (b) December 2013, (c) May 2014, and (d) January 2015.  The radial component is shown in red, azimuthal component in blue, and meridional component in green. Solid dark lines represent the result using the parameters used for mapping (inclination \ang{45}, $v \sin i~\SI{15.9}{\kilo\meter\per\second}$). Shaded areas represent varying inclination $\pm \ang{10}$ and $v \sin i \pm \SI{1}{\kilo\meter\per\second}$.  The reversal in the radial field is evident between December 2013 and May 2014.  The peak intensity of the field components occur at slightly lower latitudes as the reversal approaches, then revert to higher latitudes after the reversal.  It is noted that the above plots assume an annual magnetic field reversal (see further discussion in Section~\protect\ref{sec:conclusion}).}
  \label{fig:fracmag}
 \end{figure*}

Figure~\ref{fig:magconfig} presents the magnetic field configuration information from Tables~\ref{table:results} and \ref{table:resultsmar}, also showing the location of the approximate date of the magnetic polarity reversal. The rapid change in the activity during March through May 2015 would seem to suggest that the magnetic activity cycle is be coincident with the chromospheric activity cycle of $\tau$ Bo\"otis.  Were reversals coincident with the peak of the chromospheric activity cycle, the similarity of the 2013 though January 2015 results would appear to be more related to the similarity their position in the chromospheric cycle more than any intrinsic variation between reversals.  Figure~\ref{fig:fracmaglat} also shows that the decrease in peak fractional latitude (interpreted as solar-like behavior above) is reversed in the March-May 2015 epoch.  This is perhaps due to the strengthening of the polar features in conjunction with the overall magnetic field (Fig.~\ref{fig:magbmod}, and it would be interesting to see if the peak fractional latitude began to decrease as the chromospheric cycle continued.

Examining previous published results, no reversal between January and June 2008 would appear to settle the matter of a yearly reversal.  However, given the large gap in observations between regular observations in 2008 and 2013 and the fact that three times the observational period is not exactly a year, the approximate time of the reversal may have drifted.  It is possible two reversals may have occurred in the period between January and June 2008.  Further spectropolarimetric observations of $\tau$ Bo\"otis during each of the three $\sim\SI{117}{\day}$ chromospheric cycles would be required to confirm the periodicity of the polarity reversal.  If the reversals are confirmed to be on a yearly basis, then a mechanism whereby the 3:1 relationship between chromospheric activity and magnetic activity would need to be posited.  Magnetohydrodynamic simulations of dynamo and convective processes in F-type stars by \citet{b35} shows magnetic energy rising and falling regularly with a magnetic reversal occurring on the third such magnetic cycle thus potentially providing such a mechanism.

Our conclusions about the behaviour of $\tau$ Bo\"otis should not be significantly altered were the reversal shown to be shorter.  As our March through May 2015 observations show, the broad solar-like behaviour of the magnetic field of $\tau$ Bo\"otis should still be present, simply on a shorter timescale.

Irrespective of future work on determining the reversal cycle, an observational campaign such as March through May 2015 which spanned the maximum of the chromospheric activity would be very interesting, especially during the period of the magnetic reversal.  This would provide insights into the evolution of the magnetic field leading up to, during and following a reversal.

\citet{b26} performed observations of $\tau$ Bo\"otis in April-May 2013 which were coincident with the April-May 2013 epoch presented in this work.  Unfortunately, while they observe a plage at high latitude near $\phi_{rot}\approx 0.1$, our observations did not provide any coverage centred on that phase.  However, we do see a strong polar/high latitude magnetic feature covering  $\phi_{rot}\approx 0.85$ to $\phi_{rot}\approx 0.0$.  Given the lack of phase-coincident observations, we cannot confirm or rule out a bright spot observed by \citet{b26}.

While casual observation of the maps we present in this work seem to consistently show a feature present near $\phi_{rot}\approx 0.8$, potentially coincident with the bright spot posited by \citet{b15}, analysis of all of the data does not reliably show a persistent active longitude.  However, not every data set has coverage of this phase, and those that do may only have one or two observations over several rotations.  Thus we can neither either rule out nor confirm if there are magnetic features potentially induced by SPI or other means.

In summary, $\tau$ Bo\"otis is a prime candidate for further investigation.  It exhibits a complex interplay of chromospheric and magnetic cycles.  Whether the hot Jupiter orbiting the star is affecting these cycles is inconclusive, however further observations may be able to provide more compelling information in the future.

%
\begin{table*}
 \centering
 \begin{minipage}{130mm}
  \caption{Summary of magnetic topology evolution of $\tau$ Bo\"otis 2011-2015.  The columns indicated mean magnetic field $B$, percentage of magnetic energy in the toroidal component, percentage of the energy contained in the axisymmetric modes of the poloidal component (modes
with $m = 0$) and percentage of the energy contained in the modes of $\ell \leq 2$ of the poloidal component.  Variations are based on systematic recalculation based upon varying stellar parameters ($\Omega_{eq}$, $d\Omega$, inclination, $v \sin i$) as described in Appendix~\protect\ref{sec:appA}.}
\label{table:results}
\centering
  \begin{tabular}{@{}lccccccccc@{}}
  \hline
   Epoch & $B$ (G) & \% toroidal & \% axisymm. & \% $\ell \leq 2$ in \\
 &  &  & poloidal & poloidal \\
 \hline \\[-1.5ex]

2011 May & $2.5^{+0.4}_{-0.1}$ & $20^{+11}_{-5}$ & $19^{+14}_{-6}$ & $26^{+17}_{-7}$ \\[3pt]
2013 April/May & $2.4^{+0.6}_{-0.1}$ & $22^{+6}_{-9}$ & $29^{+8~}_{-8~}$ & $46^{+11}_{-7}$ \\[3pt]
2013 December & $3.9^{+0.4}_{-0.1}$ &  $24^{+6}_{-5}$ & $30^{+10}_{-6}$ & $40^{+12}_{-11}$ \\[3pt]
2014 May & $1.8^{+0.3}_{-0.2}$ & $13^{+6}_{-1}$ & $46^{+6~}_{-8~}$ & $45^{+12}_{-14}$ \\[3pt]
2015 Jan & $2.5^{+0.6}_{-0.2}$ & $30^{+8}_{-4}$ &  $35^{+17}_{-28}$ & $41^{+18}_{-22}$ \\[3pt]
2015 March & \multicolumn{4}{c}{see Table~\ref{table:resultsmar}} \\
\hline
\end{tabular}
\end{minipage}
\end{table*}

\begin{table*}
 \centering
 \begin{minipage}{130mm}
  \caption{Summary of magnetic topology evolution of $\tau$ Bo\"otis for the four maps March through May 2014.  The columns indicated mean magnetic field $B$, percentage of magnetic energy in the toroidal component, percentage of the energy contained in the axisymmetric modes of the poloidal component (modes with $m = 0$) and percentage of the energy contained in the modes of $\ell \leq 2$ of the poloidal component.  Variations are based on the systematic recalculation based upon varying stellar parameters ($\Omega_{eq}$, $d\Omega$, inclination, $v \sin i$) as described in Appendix~\protect\ref{sec:appA}.}
\label{table:resultsmar}
\centering
  \begin{tabular}{@{}lccccccccc@{}}
  \hline
   Epoch & $B$ (G) & \% toroidal & \% axisymm. & \% $\ell \leq 2$ in \\
 &  &  & poloidal & poloidal \\
 \hline \\[-1.5ex]

2015 02 Apr - 30 Apr & $1.2^{+0.4}_{-0.1}$ & $~5^{+6}_{-2}$ & $69^{+16}_{-9}$ & $83^{+10}_{-12}$ \\[3pt]
2015 13 Apr - 12 May & $0.9^{+0.3}_{-0.1}$ & $~7^{+1}_{-3}$ & $68^{+18}_{-8}$ & $90^{+2~}_{-6~}$ \\[3pt]
2015 20 Apr - 17 May & $1.2^{+0.2}_{-0.1}$ &  $15^{+8}_{-2}$ & $54^{+1~}_{-6~}$ & $81^{+3~}_{-6~}$ \\[3pt]
2015 12 May - 27 May & $1.9^{+0.1}_{-0.1}$ & $16^{+3}_{-3}$ & $16^{+6~}_{-5~}$ & $51^{+8~}_{-5~}$ \\[3pt]
\hline
\end{tabular}
\end{minipage}
\end{table*}

\section*{Acknowledgments}

This work was based on observations obtained with HARPSpol at the ESO 3.6-m telescope at La Silla and with NARVAL at the T\'elescope Bernard Lyot (TBL). TBL/NARVAL are operated by INSU/CNRS.  Previous spectropolarimetric observations of $\tau$ Bo\"otis were obtained from the Polarbase repository of EsPaDOns and NARVAL observations.  In particular we thank the BCool Collaboration for providing time in their long-term program for ongoing observations of $\tau$ Bo\"otis.

SVJ acknowledges research funding by the Deutsche Forschungsgemeinschaft (DFG) under grant SFB 963/1, project A16.  CPF is supported by the grant ANR 2011 Blanc SIMI5-6 020 01 ``Toupies: Towards understanding the spin evolution of stars''.

The Strategic Research Funding for the Starwinds project provided by the University of Southern Queensland provides resources to the Astrophysics group within the Computational Engineering and Science Research Centre at USQ (MWM, SCM, BDC).  MWM is supported by an Australian Postgraduate Award Scholarship.

We thank Claude Catala and Monica Rainer for their assistance in the development of the FT differential code produced by MWM for use in this paper.

Iterative calculations for differential rotation were performed using the University of Southern Queensland High Performance Computing cluster.  We thank Richard Young for his assistance in utilising the HPC resources at USQ.

This research has made use of NASA's Astrophysics Data System.

\bibliography{paper.bib}
\bibliographystyle{mnras}

\appendix

\section[]{Deriving variation measurements for magnetic field configuration}
\label{sec:appA}

As the energy in the spherical harmonics produced by the ZDI mapping process provide exact values for a given set of modes, a measure of variability of the field configuration values (Tables~\ref{table:results} and \ref{table:resultsmar}), due to the uncertainties in the stellar parameters, is desirable.  We achieve this by varying the stellar parameters, re-running the mapping process and extracting the various parameters.  As variation in $v \sin i$ and stellar inclination are used in determining the variation of the differential rotation parameters ($\Omega_{eq}$, $d\Omega$) (see Fig.~\ref{fig:drms}(b)), we hold each pair of parameters ($v \sin i$, inclination) and ($\Omega_{eq}$, $d\Omega$) constant while varying the others.  Otherwise we generate extreme variations where all parameters are varied to extremes.

\begin{figure}
\begin{center}
\includegraphics[scale=0.45]{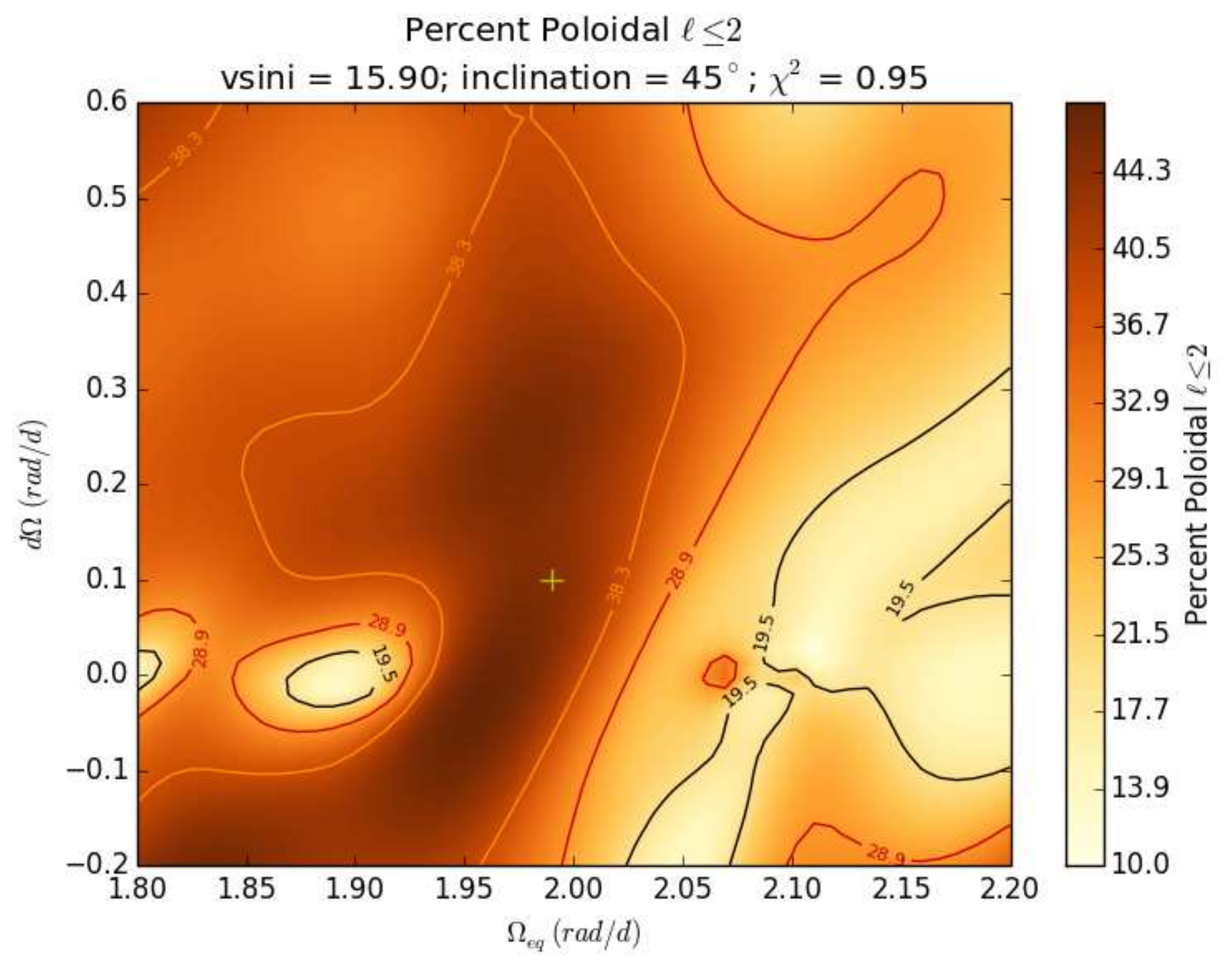}
\end{center}
\caption{Plot of percentage of poloidal modes $\ell \leq 2$ varying $\Omega_{eq}$ and $d\Omega$ for the May 2014 data set, using the values of $v \sin i = \SI{15}{\kilo\meter\per\second}$ and inclination of \ang{45}.  The yellow cross indicates the measured differential rotation.}
\label{fig:omdomvar}
\end{figure}

Figure~\ref{fig:omdomvar} shows a plot of the percentage of poloidal field modes $\ell \leq 2$ from the May 2014 data set where we vary ($\Omega_{eq}$, $d\Omega$), holding ($v \sin i$, inclination) constant.  Taking the $\sim1\mbox{-}\sigma$ variation in ($\Omega_{eq}$, $d\Omega$) from the measured value (shown as a yellow cross in Fig.~\ref{fig:omdomvar}; Tables~\ref{table:dr} and \ref{table:drmar} show variations used ) gives us an indication of how the differential rotation measurement affects this field component.

\begin{figure}
\begin{center}
\includegraphics[scale=0.45]{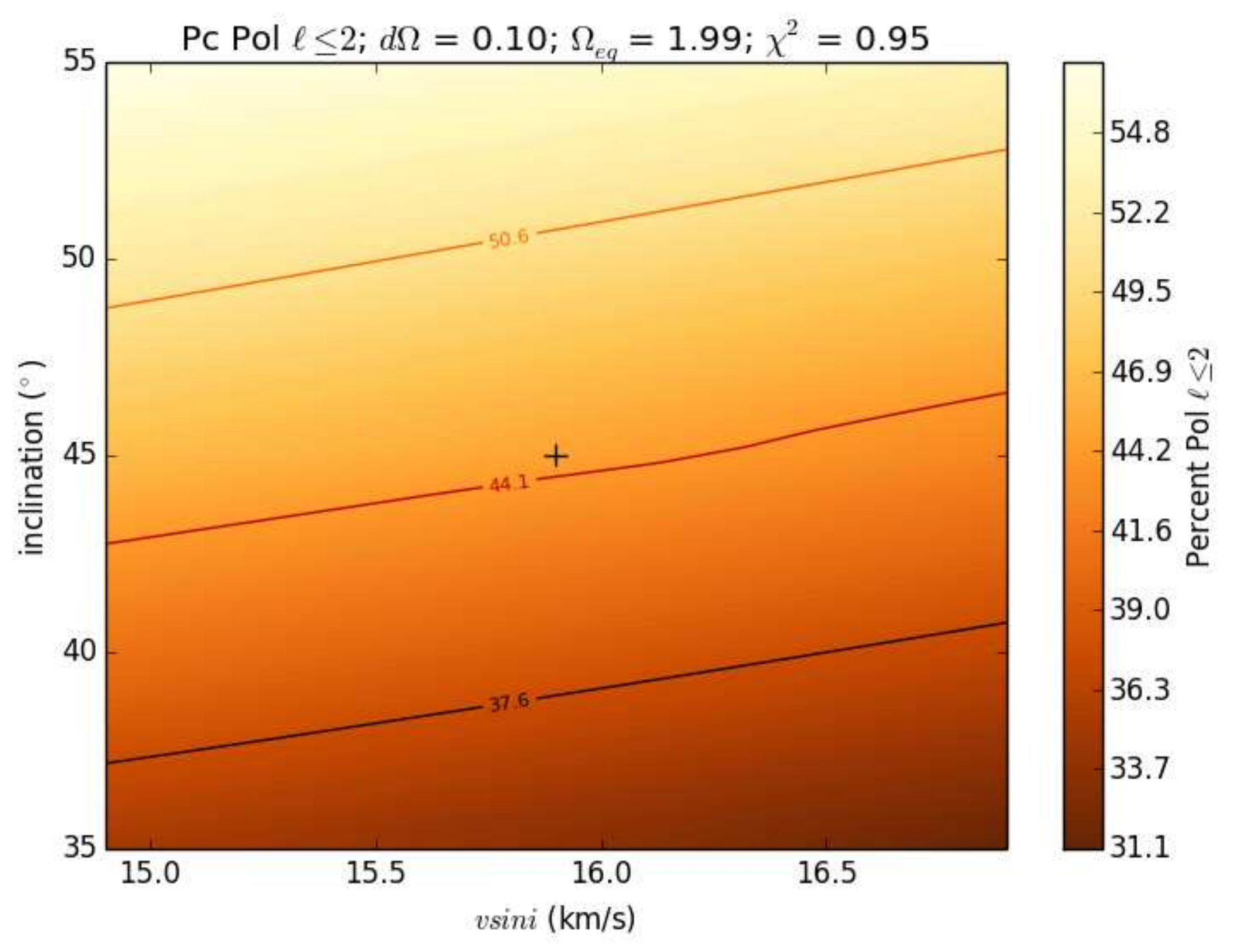}
\end{center}
\caption{Plot of percentage of poloidal modes $\ell \leq 2$ varying $v \sin i$ and angle of inclination for the May 2014 data set, using the measured values for differential rotation.  The black cross indicates the $v \sin i$ and inclination chosen for the nominal reconstruction via $\chi^{2}$ minimization described in Section~\protect\ref{sec:mapping}.}
\label{fig:vsinincvar}
\end{figure}

Figure~\ref{fig:vsinincvar} similarly shows the effect on the measured field component of varying $v \sin i$ and inclination angle while holding the DR parameters steady at the measured values.

From these measurements, we take the extreme variations from both methods to provide the variations shown in Tables~\ref{table:results} and \ref{table:resultsmar}.

It is clear that the dominant parameter for variation of poloidal axissymmetry and dipolar/quadrupolar components is the stellar inclination angle we choose.  This is as expected, as errors in the assumed inclination of the star's rotational axis will affect axissymmetry and latitudes of features.  As we use a large variation in inclination ($\pm\ang{10}$), our derived variations in these values are systematically larger than those in the percentage of toroidal field and $B_{mod}$.

Extreme changes in $\Omega_{eq}$ are significant, but usually, this value is relatively well constrained.  Extreme variation in $d\Omega$ can have an effect on the variation in the toroidal field component, but only if the variation is very large.  If variation of $d\Omega$ is not significantly greater than the $\sim1\mbox{-}\sigma$ variation in measured value, $d\Omega$ appears not to have a large effect on the field configuration parameters we derive.

In conclusion, we believe the field configuration parameters we derive are robust and stable, assuming the methodology we use is internally consistent using the derived parameters for $v \sin i$, inclination and differential rotation.

\section[]{Journals of Observations}
\label{sec:appB}

\begin{table*}
 \centering
 \begin{minipage}{160mm}
  \caption{Journal of HARPSpol observations of $\tau$ Bo\"otis. Columns list the UT date, instrument used, the heliocentric Julian date (at midpoint of the series of 4 sub-exposures),
the UT time (at midpoint of the series of 4 sub-exposures), the exposure time, the peak signal-to-noise ratio (SNR) of each observation (around \SI{583}{\nano\meter} for HARPSpol observations), the rotational cycle from the ephemeris (from Equation~\protect\ref{eq:ephem}), the rotational phase (0.0 being approximately the centre of the observing run), the radial velocity (RV) associated with each exposure, and whether a magnetic signature is detected (D; $fap < 10^{-5}$), marginally detected (M; $10^{-3} > fap > 10^{-5}$) or is below the detection threshold (N).}
\label{table:obsHARPS}
  \begin{tabular}{@{}cccccrcrcc@{}}
  \hline
   Date & Instrument & HJD & \textsc{ut} & $T_{exp}$ & SNR & Cycle & $\phi_{rot}$ & RV & Detection \\
        &  & (245 5000$+$) & (h:m:s) & (s) & & & & (\SI{}{\kilo\meter\per\second}) &   \\
 \hline
15-May-2011 & HARPSpol & 696.57356 & 01:39:37 & 4 $\times$ 600 & 1081 & 677.921 & -1.079 & -16.613 & N \\
15-May-2011 & HARPSpol & 696.65020 & 03:29:59 & 4 $\times$ 600 & 1101 & 677.945 & -1.056 & -16.544 & N \\
15-May-2011 & HARPSpol & 696.72435 & 05:16:46 & 4 $\times$ 600 & 1025 & 677.967 & -1.033 & -16.483 & N \\
16-May-2011 & HARPSpol & 697.57360 & 01:39:44 & 4 $\times$ 600 & 1074 & 678.223 & -0.777 & -15.919 & M \\
16-May-2011 & HARPSpol & 697.64667 & 03:24:57 & 4 $\times$ 600 & 1149 & 678.245 & -0.755 & -15.909 & N \\
16-May-2011 & HARPSpol & 697.72308 & 05:14:59 & 4 $\times$ 600 & 1117 & 678.268 & -0.732 & -15.914 & N \\
17-May-2011 & HARPSpol & 698.55641 & 01:15:01 & 4 $\times$ 600 & 1052 & 678.520 & -0.480 & -16.455 & M \\
17-May-2011 & HARPSpol & 698.63125 & 03:02:48 & 4 $\times$ 600 & 1065 & 678.543 & -0.457 & -16.520 & N \\
17-May-2011 & HARPSpol & 698.70306 & 04:46:12 & 4 $\times$ 600 & 1124 & 678.564 & -0.436 & -16.574 & M \\
17-May-2011 & HARPSpol & 699.50424 & 23:59:57 & 4 $\times$ 600 &  838 & 678.806 & -0.194 & -16.815 & N \\
18-May-2011 & HARPSpol & 699.57741 & 01:45:19 & 4 $\times$ 600 & 1029 & 678.828 & -0.172 & -16.792 & N \\
18-May-2011 & HARPSpol & 699.65246 & 03:33:24 & 4 $\times$ 600 & 1069 & 678.851 & -0.149 & -16.777 & D \\
19-May-2011 & HARPSpol & 700.53841 & 00:49:12 & 4 $\times$ 600 & 970 & 679.118 & 0.118 & -16.052 & N \\
19-May-2011 & HARPSpol & 700.61374 & 02:37:41 & 4 $\times$ 600 & 964 & 679.141 & 0.141 & -15.999 & N \\
19-May-2011 & HARPSpol & 700.68619 & 04:22:01 & 4 $\times$ 600 & 998 & 679.163 & 0.163 & -15.968 & N \\
20-May-2011 & HARPSpol & 701.56019 & 01:20:38 & 4 $\times$ 600 & 778 & 679.427 & 0.427 & -16.183 & N \\
20-May-2011 & HARPSpol & 701.67331 & 04:03:32 & 4 $\times$ 600 & 874 & 679.461 & 0.461 & -16.279 & N \\
20-May-2011 & HARPSpol & 701.70345 & 04:46:56 & 4 $\times$ 600 & 597 & 679.470 & 0.470 & -16.299 & N \\

\hline
\end{tabular}
\end{minipage}
\end{table*}

\begin{table*}
 \centering
 \begin{minipage}{160mm}
  \caption{Journal of NARVAL observations up to and including January 2015 of $\tau$ Bo\"otis. Columns list the UT date, instrument used, the heliocentric Julian date (at midpoint of the series of 4 sub-exposures),
the UT time (at midpoint of the series of 4 sub-exposures), the exposure time, the peak signal-to-noise ratio (SNR) of each observation (at around \SI{700}{\nano\meter} for NARVAL), the rotational cycle from the ephemeris (from Equation~\protect\ref{eq:ephem}), the rotational phase (0.0 being approximately the centre of the observing run), the radial velocity (RV) associated with each exposure, and whether a magnetic signature is detected (D; $fap < 10^{-5}$), marginally detected (M; $10^{-3} > fap > 10^{-5}$) or is below the detection threshold (N).}
\label{table:obsNARVAL}
  \begin{tabular}{@{}cccccrcrcc@{}}
  \hline
   Date & Instrument & HJD & \textsc{ut} & $T_{exp}$ & SNR & Cycle & $\phi_{rot}$ & RV & Detection \\
        &  & (245 5000$+$) & (h:m:s) & (s) & & & & (\SI{}{\kilo\meter\per\second}) &   \\
 \hline
23-Apr-2013 & NARVAL & 1405.57009 & 01:33:59 & 4 $\times$ 600 & 1493 & 891.961 & -3.040 & -16.727 & N \\
24-Apr-2013 & NARVAL & 1406.53353 & 00:41:21 & 4 $\times$ 600 & 1365 & 892.251 & -2.749 & -16.124 & D \\
24-Apr-2013 & NARVAL & 1407.41519 & 21:50:57 & 4 $\times$ 600 & 1521 & 892.518 & -2.483 & -16.631 & N \\
04-May-2013 & NARVAL & 1417.46698 & 23:05:45 & 4 $\times$ 600 & 1491 & 895.552 & 0.552 & -16.773 & M \\
05-May-2013 & NARVAL & 1418.44661 & 22:36:27 & 4 $\times$ 600 & 1410 & 895.848 & 0.848 & -16.938 & D \\
11-May-2013 & NARVAL & 1424.42736 & 22:08:58 & 4 $\times$ 600 & 1507 & 897.653 & 2.653 & -16.998 & N \\
12-May-2013 & NARVAL & 1425.42272 & 22:02:19 & 4 $\times$ 600 & 1401 & 897.954 & 2.954 & -16.669 & D \\
13-May-2013 & NARVAL & 1426.44343 & 22:32:11 & 4 $\times$ 600 & 1521 & 898.262 & 3.262 & -16.095 & D \\
& & & & & & & & & \\
04-Dec-2013 & NARVAL & 1630.73709 & 05:46:20 & 4 $\times$ 600 & 1408 & 959.936 & -3.064 & -16.822 & N \\
05-Dec-2013 & NARVAL & 1631.72401 & 05:27:25 & 4 $\times$ 600 & 1318 & 960.234 & -2.766 & -16.164 & D \\
06-Dec-2013 & NARVAL & 1632.73127 & 05:37:48 & 4 $\times$ 600 & 1280 & 960.538 & -2.462 & -16.735 & N \\
07-Dec-2013 & NARVAL & 1633.71947 & 05:20:42 & 4 $\times$ 600 & 1202 & 960.837 & -2.164 & -16.993 & M \\
08-Dec-2013 & NARVAL & 1634.71910 & 05:20:05 & 4 $\times$ 600 & 1484 & 961.138 & -1.862 & -16.268 & D \\
09-Dec-2013 & NARVAL & 1635.74031 & 05:50:32 & 4 $\times$ 600 & 1479 & 961.447 & -1.553 & -16.444 & D \\
11-Dec-2013 & NARVAL & 1637.72461 & 05:27:44 & 4 $\times$ 600 & 1259 & 962.046 & -0.954 & -16.520 & M \\
12-Dec-2013 & NARVAL & 1638.71993 & 05:20:54 & 4 $\times$ 600 & 1546 & 962.346 & -0.654 & -16.257 & D \\
13-Dec-2013 & NARVAL & 1639.73536 & 05:43:02 & 4 $\times$ 600 & 1505 & 962.653 & -0.347 & -17.022 & D \\
15-Dec-2013 & NARVAL & 1641.72308 & 05:25:09 & 4 $\times$ 600 & 1381 & 963.253 & 0.253 & -16.116 & N \\
17-Dec-2013 & NARVAL & 1643.75082 & 06:04:53 & 4 $\times$ 600 & 1483 & 963.865 & 0.865 & -16.923 & D \\
21-Dec-2013 & NARVAL & 1647.73550 & 05:42:25 & 4 $\times$ 600 & 1447 & 965.068 & 2.068 & -16.403 & N \\
& & & & & & & & & \\
04-May-2014 & NARVAL & 1782.47281 & 23:14:08 & 4 $\times$ 600 & 1304 & 1005.744 & -2.256 & -17.089 & D \\
05-May-2014 & NARVAL & 1783.47546 & 23:17:58 & 4 $\times$ 600 & 1350 & 1006.046 & -1.954 & -16.397 & N \\
07-May-2014 & NARVAL & 1785.46095 & 22:57:09 & 4 $\times$ 600 & 1079 & 1006.646 & -1.354 & -16.945 & M \\
08-May-2014 & NARVAL & 1786.49026 & 23:39:24 & 4 $\times$ 600 & 1377 & 1006.957 & -1.044 & -16.668 & M \\
09-May-2014 & NARVAL & 1787.48744 & 23:35:22 & 4 $\times$ 600 & 1374 & 1007.258 & -0.742 & -16.094 & D \\
14-May-2014 & NARVAL & 1791.54389 & 00:56:51 & 4 $\times$ 600 & 1512 & 1008.482 & 0.482 & -16.563 & N \\
14-May-2014 & NARVAL & 1792.43744 & 22:23:36 & 4 $\times$ 600 & 954 & 1008.752 & 0.752 & -17.089 & N \\
15-May-2014 & NARVAL & 1793.48608 & 23:33:42 & 4 $\times$ 600 & 940 & 1009.069 & 1.069 & -16.438 & N \\
16-May-2014 & NARVAL & 1794.46699 & 23:06:15 & 4 $\times$ 600 & 892 & 1009.365 & 1.365 & -16.345 & N \\
17-May-2014 & NARVAL & 1795.47600 & 23:19:17 & 4 $\times$ 600 & 1312 & 1009.669 & 1.669 & -17.048 & M \\
18-May-2014 & NARVAL & 1796.45889 & 22:54:42 & 4 $\times$ 600 & 1295 & 1009.966 & 1.966 & -16.671 & M \\
& & & & & & & & & \\
06-Jan-2015 & NARVAL & 2028.73834 & 05:44:44 & 4 $\times$ 600 & 1511 & 1080.089 & -1.911 & -16.355 & D \\
07-Jan-2015 & NARVAL & 2029.72424 & 05:24:19 & 4 $\times$ 520 & 1386 & 1080.387 & -1.614 & -16.372 & D \\
08-Jan-2015 & NARVAL & 2030.74613 & 05:55:43 & 4 $\times$ 520 & 1401 & 1080.695 & -1.305 & -17.149 & N \\
09-Jan-2015 & NARVAL & 2031.72258 & 05:21:41 & 4 $\times$ 600 & 1396 & 1080.990 & -1.010 & -16.673 & D \\
10-Jan-2015 & NARVAL & 2032.73855 & 05:44:33 & 4 $\times$ 520 & 1442 & 1081.297 & -0.704 & -16.207 & M \\
12-Jan-2015 & NARVAL & 2034.73935 & 05:45:29 & 4 $\times$ 520 & 1425 & 1081.901 & -0.100 & -16.988 & D \\
18-Jan-2015 & NARVAL & 2040.67474 & 04:11:43 & 4 $\times$ 600 & 707 & 1083.692 & 1.692 & -17.178 & N \\

\hline
\end{tabular}
\end{minipage}
\end{table*}

\begin{table*}
 \centering
 \begin{minipage}{160mm}
  \caption{Journal of NARVAL observations up to and including March through May 2015 of $\tau$ Bo\"otis. Columns list the UT date, instrument used, the heliocentric Julian date (at midpoint of the series of 4 sub-exposures),
the UT time (at midpoint of the series of 4 sub-exposures), the exposure time, the peak signal-to-noise ratio (SNR) of each observation (at around \SI{700}{\nano\meter} for NARVAL), the rotational cycle from the ephemeris (from Equation~\protect\ref{eq:ephem}), the radial velocity (RV) associated with each exposure, and whether a magnetic signature is detected (D; $fap < 10^{-5}$), marginally detected (M; $10^{-3} > fap > 10^{-5}$) or is below the detection threshold (N).}
\label{table:obsNARVAL2015}
  \begin{tabular}{@{}cccccrccc@{}}
  \hline
   Date & Instrument & HJD & \textsc{ut} & $T_{exp}$ & SNR & Cycle & RV & Detection \\
        &  & (245 5000$+$) & (h:m:s) & (s) & & & (\SI{}{\kilo\meter\per\second}) &   \\
 \hline
12-Mar-2015 & NARVAL & 2093.60723 & 02:28:57 & 4 $\times$ 600 & 1300 & 1099.672 & -17.103 & N \\
13-Mar-2015 & NARVAL & 2094.57542 & 01:43:04 & 4 $\times$ 600 & 1387 & 1099.964 & -16.769 & M \\
17-Mar-2015 & NARVAL & 2098.60339 & 02:23:04 & 4 $\times$ 600 & 1243 & 1101.180 & -16.307 & D \\
02-Apr-2015 & NARVAL & 2114.54024 & 00:51:19 & 4 $\times$ 600 & 1567 & 1105.992 & -16.685 & M \\
13-Apr-2015 & NARVAL & 2126.44800 & 22:38:12 & 4 $\times$ 600 & 1357 & 1109.586 & -17.002 & M \\
14-Apr-2015 & NARVAL & 2127.37286 & 20:49:59 & 4 $\times$ 600 &  592 & 1109.866 & -17.016 & N \\
20-Apr-2015 & NARVAL & 2133.42700 & 22:07:56 & 4 $\times$ 600 & 1369 & 1111.693 & -17.159 & M \\
21-Apr-2015 & NARVAL & 2134.42323 & 22:02:30 & 4 $\times$ 600 &  590 & 1111.994 & -16.650 & N \\
23-Apr-2015 & NARVAL & 2136.45663 & 22:50:36 & 4 $\times$ 600 & 1446 & 1112.608 & -17.010 & D \\
30-Apr-2015 & NARVAL & 2142.52088 & 00:23:13 & 4 $\times$ 600 & 1385 & 1114.439 & -16.554 & N \\
11-May-2015 & NARVAL & 2154.47299 & 23:14:38 & 4 $\times$ 600 &  992 & 1118.047 & -16.542 & N \\
12-May-2015 & NARVAL & 2155.47975 & 23:24:25 & 4 $\times$ 600 & 1043 & 1118.351 & -16.235 & N \\
16-May-2015 & NARVAL & 2159.45312 & 22:46:16 & 4 $\times$ 600 & 1417 & 1119.550 & -16.848 & M \\
17-May-2015 & NARVAL & 2160.48669 & 23:34:39 & 4 $\times$ 600 & 1077 & 1119.862 & -17.049 & N \\
19-May-2015 & NARVAL & 2161.50449 & 00:00:21 & 4 $\times$ 700 &  472 & 1120.170 & -16.243 & N \\
27-May-2015 & NARVAL & 2170.37187 & 20:49:57 & 4 $\times$ 600 & 1527 & 1122.847 & -17.025 & D \\
\hline
\end{tabular}
\end{minipage}
\end{table*}

\label{lastpage}

\end{document}